\documentclass[12pt]{article}    
      
\usepackage{setspace,url} 
\usepackage{dsfont,kz_style}   
\usepackage{ifthen}
\usepackage{mathtools}
\usepackage{graphicx,epstopdf}
\usepackage{enumerate,comment}
\usepackage{enumitem}
\usepackage{color,xcolor,soul}
\usepackage[margin=1.0in]{geometry} 
\usepackage{natbib}

\usepackage{algorithm}
\usepackage{algpseudocode}
\usepackage[capitalize]{cleveref}
\usepackage{palatino}
\usepackage{cuted,balance} 
\usepackage{cancel,array}

\newcommand{\player}[1]{player $#1$}

\newcommand{\val}{\mathrm{val}}

\newcommand{\smooth}{\overline{\mathrm{BR}}}

\usepackage{tikz}  

\begin{document}

\date{}

\title{Independent Learning in Stochastic Games} 
\author{Authors}
\author{Asuman Ozdaglar\thanks{Invited chapter for  International Congress of Mathematicians, 2022. A. Ozdaglar and K. Zhang are with the Department of Electrical Engineering and Computer Science, Massachusetts Institute of Technology, Cambridge MA. E-mail: \texttt{\{asuman,kaiqing\}@mit.edu}.} \and Muhammed O. Sayin\thanks{M. O. Sayin is with the Electrical and Electronics Engineering Department in Bilkent University, Turkey. E-mail: \texttt{sayin@ee.bilkent.edu.tr}.} \and Kaiqing Zhang\footnotemark[1]}
\maketitle 
\thispagestyle{empty}   

\bigskip
 
\begin{abstract}
Reinforcement learning (RL) has recently achieved tremendous successes in many  artificial intelligence applications. 
Many of the forefront applications of RL involve   \emph{multiple agents}, e.g., playing chess and Go games, autonomous driving, and robotics. Unfortunately, the framework upon which classical RL builds is inappropriate for multi-agent learning, as it assumes an agent's environment is stationary and does not take into account the adaptivity of other agents.  In this review paper, we present the model of \emph{stochastic games} \citep{ref:Shapley53} for multi-agent learning  in \emph{dynamic}  environments. We focus on the development of \emph{simple} and \emph{independent}  learning dynamics for stochastic games: each agent is myopic and chooses best-response type  actions to other agents' strategy without any coordination with her opponent. There has been limited progress on developing convergent best-response type independent learning dynamics for stochastic games. 
  We present our recently proposed simple and independent learning dynamics  that guarantee convergence in zero-sum stochastic games, together with a  review of other contemporaneous  algorithms  for dynamic multi-agent learning in this setting. Along the way, we also reexamine some classical results from both the game theory and RL literature, to situate both the conceptual contributions of our independent learning dynamics, and the mathematical novelties of our analysis.  
  We hope this review paper serves as an impetus for the resurgence of  studying independent and natural learning dynamics in game theory, for the more challenging settings with a dynamic environment. 
\end{abstract}

\bigskip



\bigskip 

\bigskip

\setcounter{page}{1} 

\begin{spacing}{1.245}


\section{Introduction}

Reinforcement learning (RL) in which autonomous agents make decisions in unknown dynamic environments has emerged as the backbone of many artificial intelligence (AI) problems. The frontier of many AI systems emerges in \emph{multi-agent} settings, including playing  games such as chess and Go \citep{silver2016mastering, silver2017mastering}, robotic manipulation with multiple connected arms \citep{gu2017deep}, autonomous vehicle control in dynamic traffic and automated warehouses or production facilities \citep{shalev2016safe,yang2020multi}. Further advances in these problems critically  depend  on developing stable and agent incentive-compatible  learning dynamics in multi-agent environment.  Unfortunately, the mathematical framework upon which classical RL depends on is inadequate for  multi-agent learning,  since it assumes an agent's environment is stationary and does not contain any adaptive agents. 

The topic of multi-agent learning has a long history in game theory, almost as long as the discipline itself. One of the most studied models of learning in games is \emph{fictitious play}, introduced by Brown \citep{brown1951iterative}, with first rigorous convergence analysis presented by Robinson \citep{robinson1951iterative} for its discrete-time variant and for finite two-player  zero-sum games. See also \cite{ref:Miyasawa61,shapley1964some,fudenberg1993learning,fudenberg1995consistency,ref:Monderer96b,hofbauer2002global} and others for the analysis of fictitious play. In fictitious play, each agent is myopic (i.e., she does not take into account the fact that her current action will have an impact  on the future actions of other players\footnote{Hereafter, we use \emph{player} and \emph{agent}  interchangeably.}), and therefore chooses a best response to the opponent's strategy, which she estimates to be the empirical distribution of past play. Despite extensive study on learning in repeated play of static complete-information games (also referred to as strategic-form or normal-form games) and the importance of the issues, there is limited progress on multi-agent learning in dynamic environments (where the environments evolve over time). The key challenge is to estimate the decision rules of other agents that in turn \emph{adapt} their behavior to changing non-stationary environments.  

In this review paper, we first present stochastic games, first introduced by \cite{ref:Shapley53}, as a model for representing dynamic multi-agent interactions in \S\ref{sec:stochastic}.\footnote{The preliminary information on strategic-form games and learning in strategic-form games with repeated play are provided in \S\ref{sec:strategic-form}.} Stochastic games extend strategic-form games to dynamic settings where the environment changes with players' 
 decisions. They also extend single-agent Markov decision problems (Markov decision processes) to competitive situations with more than one decision-maker.  Developing simple and independent learning rules, e.g., the fictitious-play/best-response  type  dynamics,  for stochastic games has been an open question for some time in the literature (see \cite{condon1990algorithms,tan1993multi,claus1998dynamics} for some negative non-convergent results due to non-stationarity).  
  
In the second part of the paper in \S\ref{sec:learning}, we present recently proposed simple and independent learning rules from \citep{ref:Sayin20,ref:Sayin21}, and show their convergence for zero-sum stochastic games. Crucially, these rules are based on fictitious play-type dynamics and unlike earlier works, do not require coordination between agents, leading to fully decentralized and independent multi-agent learning dynamics. We combine ideas from {\it game theory} and {\it RL}  in developing these learning rules, and consider three different settings: \emph{model-based}  setting where players know their payoff functions, transition probabilities of the underlying stochastic games, and observe opponent's actions; \emph{model-free} setting where players do not know payoff functions and transition probabilities but can still observe the opponent's actions; and the \emph{minimal information} setting where players do not even observe opponent's actions. In all three settings, the players do not know the opponent's  objective, i.e.,  they do not possess the knowledge that the underlying game is zero-sum. In the minimal-information setting, the players may not even know the existence of an opponent. 

In \S\ref{sec:other}, we have also reviewed several other  algorithms/learning dynamics, and their convergence results for multi-agent learning in stochastic games.  We cover both results from the {game theory}  literature that typically assumes knowledge of the model of the players' payoff functions, and the transition probabilities of the underlying stochastic games, and also from the {RL} literature which posit learning dynamics that perform updates without knowing the transition probabilities. Most of these  update rules typically 
involve coordination and computationally intensive steps for the players. These algorithms can be viewed more as ones for \emph{computing}  the Nash equilibrium of the stochastic games, as opposed to natural learning dynamics that would be adopted by self-interested agents interested in maximizing their own payoffs given their inferences (as captured in our learning dynamics). Finally, we conclude the paper with open questions on independent learning in stochastic games in \S\ref{sec:conclusion}. 
 
%


\section{Preliminary: Strategic-Form Games}\label{sec:strategic-form}

A two-player strategic-form game can be characterized by a tuple $\langle A^1,A^2, r^1,r^2 \rangle$, in which 
\begin{itemize}
\item The \textit{finite} set of actions that player $i$ can take is denoted by $A^i$,
\item The \textit{payoff function} of player $i$ is denoted by $r^i:A \rightarrow \mathbb{R}$, where $A:=A^1\times A^2$.\footnote{We can generalize the definition to arbitrary number of players in a rather straight-forward way.}
\end{itemize}
Each player $i$ takes an action from her  action set $A^i$  \textit{simultaneously} and receives  the payoff $r^i(a^1,a^2)$. 

We let players choose a mixed strategy to randomize their actions independently. For example, $\pi^i:A^i\rightarrow [0,1]$ denotes the mixed strategy of player $i$ such that $\pi^i(a^i)$ corresponds to the probability that player $i$ plays $a^i$. Note that we have $\sum_{a^i\in A^i} \pi^i(a^i) = 1$ by its definition. 

We represent the strategy profile and action profile of the players by $\pi=(\pi^1,\pi^2)$ and $a = (a^1,a^2)$, respectively. Under the strategy profile $\pi$, the expected payoff of player $i$ is defined by
$$
U^i(\pi) := \EE_{a\sim\pi}\left\{r^i(a)\right\}. 
$$
Note that the expected payoff of player $i$ is affected by the strategy of the opponent. We next introduce the Nash equilibrium where players do not have any (or large enough) incentive to change their strategies unilaterally.

\begin{definition}[($\varepsilon$-)Nash Equilibrium]
A strategy profile $\pi_*$ is a mixed-strategy $\varepsilon$-Nash equilibrium with $\varepsilon \geq 0$ if we have
\begin{subequations}\label{eq:NE}
\begin{align}
&U^1(\pi_*^1,\pi_*^2) \geq U^1(\pi^1,\pi_*^2) - \varepsilon, \quad \textup{for all}~~  \pi^1\\
&U^2(\pi_*^1,\pi_*^2) \geq U^2(\pi_*^1,\pi^2) - \varepsilon, \quad \textup{for all}~~  \pi^2.
\end{align}
\end{subequations}
Furthermore, $\pi_*$ is a mixed-strategy Nash equilibrium if \eqref{eq:NE} holds with $\varepsilon=0$. 
\end{definition}

The following is the classical existence result for any strategic-form game (e.g., see \citep[Theorem 3.2]{ref:Basar99}).

\begin{theorem}[Existence of An Equilibrium in Strategic-Form Games]\label{thm:NE}~~In strategic-form games (with finitely many players and finitely many actions), a mixed-strategy equilibrium always exists.
\end{theorem}

The key question is whether an equilibrium can be realized or not in the interaction of self-interested  decision-makers. In general, finding the best strategy against another  decision-maker is not a well-defined optimization problem because the best strategy that reflects the viewpoint of the individual depends  on the opponent's strategy. Therefore, players are generally not able to compute their best strategy beforehand. When there exists a unique equilibrium, we can expect the players to identify their equilibrium strategies as a result of an introspective thinking process. For example, what would the opponent choose? What would the opponent have chosen if she knew I am considering what she would pick while choosing my strategy? And so on. However, many empirical analyses suggest that an equilibrium would not typically  be realized in one shot   even with such reasoning (see, e.g., \cite{fudenberg1998theory}).

It is instructive to consider  the following well-known example: Consider a game played among $n>1$ students. The teacher asks the students to pick a number between $0$ and $100$, and submit it within a closed envelope. The winner will be the one who chooses the number closest to the $2/3$rd of the average of all numbers picked. It can be seen that the unique  equilibrium is the strategy profile where every player chooses $0$. We would expect the students to pick $0$ as a result of an introspective thinking process, however, empirical studies show that they typically  pick numbers other than zero such that their average end up around $30$ with its $2/3$rd  around $20$ \citep{ref:Nagel95}. This results in  players who have selected $0$ by strategizing their actions introspectively losing the game. However, if the game is played repeatedly with players observing chosen actions,   each player will have a tendency to pick numbers closer to the winning number (or its $2/3$rd if they notice that others can also have such a tendency to pick the number closest to the winning one). This results in convergence to the equilibrium play along repeated play of the game,  even when the players have not engaged in any forward-looking strategy. 

Many games have multiple equilibria which makes coordination and selection through introspective thinking challenging. On the other hand, empirical studies suggest even in strategic situations equipped with multiple equilibria, individual agents reach an equilibrium as long as they \textit{engage with each other multiple times and receive feedback} to revise their strategies \citep{fudenberg1998theory}. 

In the following, we review the canonical  models of learning with multiple agents through repeated interactions. 

\subsection{Learning in Strategic-Form Games with Repeated Play}\label{sec:FP}

Suppose that players know the primitives of the game, i.e., $\langle A^1,A^2, r^1,r^2 \rangle$. If players knew the opponent's strategy, computation of the best strategy is a simple optimization problem where they pick one of the maxima among linearly ordered finitely many elements. However, players do not know the opponent's strategy. When they play the same game repeatedly and observe the opponent's actions in these games, they have a chance to reason about what the opponent would play in the next repetition of the game. Therefore, they can estimate the opponent's strategy based on the history of the play. However, the opponent is not necessarily playing according to a stationary strategy since she is also a strategic decision-maker who can adapt her strategy according to her best interest. 

\textit{Fictitious play} is a simple and stylist learning dynamic where players (erroneously) assume that the opponent plays according to a stationary strategy.\footnote{It is called \textit{fictitious play} because \cite{brown1951iterative} introduced it as an introspective thinking process that a player can play by herself.} This assumption lets players form a belief on the opponent's strategy based on the history of the play, e.g., the empirical distribution of the actions taken. Then, the players can adapt their strategies based on the belief constructed.

Fictitious play, since its first introduction by \cite{brown1951iterative}, has become the most appealing  best-response type learning dynamics in game theory. 
Formally, at iteration $k$, player $i$ maintains a \emph{belief} on the opponent's strategy, denoted by $\hat{\pi}^{-i}_k\in\Delta(A^{-i})$.\footnote{We represent the probability simplex over a set $A$ by $\Delta(A)$.} For example, the belief can correspond to the empirical average of the actions taken in the past. Note that we can view an action $a^i$ as a deterministic strategy in which the action is played with probability $1$, i.e., $a^i \in \Delta(A^i)$ with slight abuse of notation. Then, the empirical average is given by
\begin{align}
\hat{\pi}_{k}^{-i} = \frac{1}{k+1}\sum_{\kappa=0}^k a_{\kappa}^{-i}. 
\end{align}
The belief $\hat{\pi}_{k+1}^{-i}$ can be computed iteratively using bounded memory according to 
\begin{align}\label{equ:FP_1}
\hat{\pi}_{k+1}^{-i} = \hat{\pi}_{k}^{-i} + \frac{1}{k+1}\cdot(a_k^{-i} - \hat{\pi}_{k}^{-i}), 
\end{align}
with arbitrary initialization $\hat{\pi}_0^{-i} \in \Delta(A^{-i})$. In other words, players do not have to remember every action taken by the opponent in the past. Moreover, player $i$ selects her action following
\begin{align}\label{equ:FP_2}
a_{k}^i \in \argmax_{a^i\in A^i} \; \mathbb{E}_{a^{-i}\sim \hat{\pi}^{-i}_k} \left\{r^i(a^1,a^2)\right\},
\end{align}
with an arbitrary tie-breaking rule, playing a greedy best-response to the belief she maintains on opponent's strategy. 

We say that fictitious play dynamics {\it converge to an equilibrium} if beliefs formed converge to a Nash equilibrium when all players follow the fictitious play dynamics \eqref{equ:FP_1}-\eqref{equ:FP_2}. We also say that a class of games has \textit{fictitious play property} if fictitious play converges in every game of that class. The following theorem is about two important classes of games from two extremes of the game spectrum: two-player zero-sum strategic-form games, where $r^1(a) + r^2(a) = 0$ for all $a\in A$, and $n$-player identical-interest strategic-form games, where there exists a common payoff function $r:A\rightarrow \mathbb{R}$ such that $r^i(a)=r(a)$ for all $a\in A$ and for each player $i$.

\begin{theorem}[Fictitious Play Property of Zero-Sum and Identical-Interest Games]~{}
\begin{itemize}
\item The two-player zero-sum strategic-form games have fictitious play property \citep{robinson1951iterative}. 
\item The $n$-player identical-interest strategic-form games have fictitious play property \citep{ref:Monderer96b}. 
\end{itemize}
\end{theorem}

Alternative to the insightful proofs in \citep{robinson1951iterative} and \citep{ref:Monderer96b}, we can establish a connection between fictitious play and continuous-time best response dynamics to characterize its convergence properties. For example, \cite{ref:Harris98} provided a proof for the continuous-time best-response dynamics in zero-sum strategic-form games through a Lyapunov function formulation. This convergence result also implies the convergence of fictitious play in repeated play of the same zero-sum strategic-form game.  We next briefly describe \cite{ref:Harris98}'s approach to convergence analysis for continuous-time best-response dynamics. 

In continuous-time best response dynamics, the strategies $(\pi^1,\pi^2)$ evolve according to the following differential inclusion
\begin{align}\label{eq:BR}
\frac{d\pi^i}{dt} + \pi^i \in \argmax_{a^i\in A^i} \;\mathbb{E}_{a^{-i}\sim \pi^{-i}} \{r^i(a^1,a^2)\}
\end{align}
for $i=1,2$. We highlight the resemblance between \eqref{equ:FP_1} and \eqref{eq:BR} because we can view \eqref{eq:BR} as the limiting flow of \eqref{equ:FP_1} as $1/(k+1)\rightarrow 0$. Note also that there exists an absolutely continuous solution to this differential inclusion \citep{ref:Harris98}. To characterize the convergence properties of this flow, \cite{ref:Harris98} showed that the function
\begin{align}\label{eq:Lyapunov}
V(\pi) = \sum_{i=1,2} \left(\max_{a^i\in A^i} \;\mathbb{E}_{a^{-i}\sim \pi^{-i}} \{r^i(a^1,a^2)\} - \mathbb{E}_{a\sim \pi} \{r^i(a)\}\right)
\end{align}
is a Lyapunov function when $r^1(a)+r^2(a)=0$ for all $a\in A$.\footnote{Note that $\mathbb{E}_{a\sim \pi} \{r^1(a)\} +  \mathbb{E}_{a\sim \pi} \{r^2(a)\} = 0$ when $r^1(a)+r^2(a)=0$ for all $a\in A$.} This yields that $V(\pi(t))\geq V(\pi(t'))$ for all $t'>t$ and $V(\pi(t)) >V(\pi(t'))$ if $V(\pi(t)) > 0$. Correspondingly, we have $V(\pi(t))\rightarrow 0$ as $t\rightarrow \infty$. This implies that the continuous-time best response dynamics converge to the equilibrium of the zero-sum game. Since the terms in parentheses in \eqref{eq:Lyapunov} are non-negative, $V(\pi)=0$ yields that they are equal to zero for each $i=1,2$, which is indeed the definition of the Nash equilibrium. 

Generally, the convergence of the limiting flow would not lead to the convergence of the discrete-time update. However, based on tools from differential  inclusion approximation theory \citep{ref:Benaim05}, the existence of such a Lyapunov function yields that the fictitious play dynamics converge to an equilibrium since its linear interpolation after certain transformation of the time axis can be viewed as a perturbed solution to the differential inclusion \eqref{eq:BR} with asymptotically negligible perturbation while the existence of Lyapunov function yields that any such perturbed solution also converges to the zero-set of the Lyapunov function, i.e., $\{\pi:V(\pi)=0\}$.

The fictitious play dynamics enjoy the following desired properties \citep{fudenberg1998theory}: $i)$ The dynamics do not require knowledge of the underlying game's class, e.g., the opponent's payoff function, and is not specific to any specific class of games; $ii)$ Players attain the best-response performance against an opponent following an asymptotically stationary strategy, i.e., the learning dynamics is rational; $iii)$ If the dynamics converge, it must converge to an equilibrium of the underlying game. 

Unfortunately, there exist strategic-form games that do not have fictitious play property as shown by \cite{shapley1964some} through a counter-example. The classes of strategic-form games with fictitious play property have been studied extensively, e.g., see \citep{robinson1951iterative,ref:Miyasawa61,ref:Milgrom91,ref:Monderer96a,ref:Monderer96b,ref:Monderer96c,ref:Sela99,ref:Berger05,ref:Berger08}. Variants of fictitious play, including smoothed fictitious play \citep{fudenberg1993learning} and weakened fictitious play \citep{ref:Genugten00} have also been studied extensively. However, all these studies focus on the repeated play of \textit{the same} strategic-form game at every stage. There are very limited results on dynamic games where players interact repeatedly while the game played at a stage (called \textit{stage-game}) evolves with their actions. Note that players need to consider the impact of their actions in their future payoffs as in dynamic programming or optimal control when they have utilities defined over infinite horizon.

In the next section, we introduce stochastic games, a special (and important class) of dynamic games where the stage-games evolve over infinite horizon based on the current actions of players. 


\section{Stochastic Games}\label{sec:stochastic}

Stochastic games (also known as \textit{Markov} games), since its first introduction by  \cite{ref:Shapley53}, have been widely used as a canonical  model for dynamic multi-agent interactions (e.g., see the surveys  \citep{busoniu2008comprehensive,ref:Zhang21}). At each time $k=0,1,\ldots$, players play a stage game that corresponds to a particular state of a multi-state environment. The stage games evolve stochastically according to the transition probabilities of the states controlled jointly by the actions of both players. The players receive a payoff which is some aggregate of the stage payoffs; a typical model is to assume the players receive a discounted sum of stage payoffs over an infinite horizon.  

Formally, a two-player stochastic game is characterized by a tuple $\langle  S , A^1,A^2,r^1,r^2,p, \gamma \rangle$, in which 
\begin{itemize}
\item The \textit{finite} set of states is denoted by $S$,
\item The \textit{finite} set of actions that player $i$ can take at any state is denoted by $A^i$,\footnote{The formulation can be generalized to the case where the action spaces depend on state in a rather straightforward way.}
\item The \textit{stage payoff function} of player $i$ is denoted by $r^{i}:S\times A\rightarrow \mathbb{R}$, where $A=A^1\times A^2$.  
\item For any pair of states $(s,s')$ and action profile $a\in A$, we define $p(s'|s,a)$ as the {\em transition probability} from $s$ to $s'$ given action profile $a$.
\item The players also discount the impact of future payoff in their utility with the discount factor $\gamma \in[0,1)$.
\end{itemize}
The objective of player $i$ is to maximize the expected sum of discounted stage-payoffs collected over infinite horizon, given by
 \begin{equation}
\mathbb{E}\left\{\sum_{k=0}^{\infty}\gamma^k r^{i}(s_k,a_{k})\right\},
 \end{equation}
where $a_k\in A$ denotes the action profile played at stage $k$, $\{s_0\sim p_o, s_{k+1}\sim p(\cdot|s_{k},a_{k}),k\ge 0\}$ is a stochastic process representing the state at each stage $k$ and $p_o\in \Delta(S)$ is the initial state distribution. The expectation is taken with respect to randomness due to stochastic state transitions and actions mixed independently by the players. 

The players can play an \textit{infinite} sequence of (mixed) actions. When they have perfect recall, they can mix their actions independently according to a \textit{behavioral strategy} in which the probability of an action is taken depends on the history of states and action profiles, e.g., $h_k = \{s_0,a_0,s_1,a_1,\ldots,s_{k-1},a_{k-1},s_k\}$ at stage $k$. This results in an infinite-dimensional strategy space, and therefore, the universal result for the existence of an equilibrium, Theorem \ref{thm:NE}, does not apply here. On the other hand, stochastic games can also be viewed as a generalization of Markov decision processes (MDPs) to multi-agent cases since state transition probabilities depend only on the current state and current action profile of players. Behavioral strategies that depend only on the final state of the history (which corresponds to the current state) are known as \textit{Markov strategies}. Furthermore, we call a Markov strategy by a \textit{stationary strategy} if it does not depend on the stage, e.g., see \citep[Section 6.2]{ref:MAS}. In (discounted) MDPs, there always exists an optimal strategy that is stationary, e.g., see \citep{filar2012competitive}. \cite{ref:Shapley53} showed that this can be generalized to two-player zero-sum stochastic games. 



We denote the stationary mixed strategy of player $i$ by $\pi^i: S\rightarrow \Delta(A^i)$, implying that she takes actions according to the mixed strategy specific to state $s$, i.e., $\pi^i(s)\in \Delta(A^i)$. We represent the strategy profile of players by $\pi:=\{\pi^1,\pi^2\}$. Correspondingly, the expected discounted sum of stage payoffs of player $i$ under the strategy profile $\pi$ is defined by 
\begin{align}\label{eq:utility}
U^i(\pi) := \mathbb{E}\left\{\sum_{k=0}^{\infty}\gamma^k r^{i}(s_k,a_{k})\right\},
\end{align}
where $a_k\sim \pi(s_k)$, and the expectation is taken with respect to the all randomness. We next introduce the Nash equilibrium (more specifically Markov perfect equilibrium \citep{maskin1988theory,maskin1988theoryb}) where players do not gain any utility  improvement by unilateral changes in their \textit{stationary} strategies regardless of the initial state, e.g., see \citep[Section 6.2]{ref:MAS}.

\begin{definition}[Stationary ($\varepsilon$-)Nash Equilibrium]
We say that a stationary strategy profile $\pi$ is a stationary mixed-strategy $\varepsilon$-Nash equilibrium with $\varepsilon\geq 0$ if we have  
\begin{subequations}\label{eq:equilibrium}
\begin{align}
&U^1(\pi^1,\pi^{2})\ge U^1(\bar{\pi}^1,\pi^{2}) - \varepsilon \quad \textup{for all}~~ \bar{\pi}^1, \\ 
&U^2(\pi^1,\pi^{2})\ge U^2(\pi^1,\bar{\pi}^2) - \varepsilon  \quad \textup{for all}~~ \bar{\pi}^2.
\end{align}
\end{subequations} 
We say that $\pi$ is a stationary mixed-strategy Nash equilibrium if \eqref{eq:equilibrium} holds with $\varepsilon = 0$. 
\end{definition}

We next state an important existence result for discounted stochastic  games. 


\begin{theorem}[Existence of A Stationary Equilibrium in Stochastic Games \citep{ref:Fink64}]~In stochastic games (with finitely many players, states and actions, and discount factor $\gamma\in[0,1)$), a stationary mixed-strategy equilibrium always exists.
\end{theorem}

The proof for two-player zero-sum stochastic games is shown by \citep{ref:Shapley53} while its generalization to $n$-player general-sum stochastic games is proven by \citep{ref:Fink64} and \citep{ref:Takahashi64} concurrently. \cite{ref:Shapley53} had also presented an iterative algorithm to compute the unique equilibrium value of a two-player zero-sum stochastic game. To describe the algorithm, let's first note that in a zero-sum strategic-form game, there always exists a {unique} equilibrium {\it value} for the players (though there may exist multiple equilibria). For example, given a zero-sum strategic-form game $\langle A^1,A^2,u^1,u^2\rangle$, we denote the equilibrium values of \player{1} and \player{2}, respectively, by
\begin{align}
&\mathrm{val}^1[u^1] = \max_{\pi^1\in\Delta(A^1)}\min_{\pi^2\in\Delta(A^2)} \mathbb{E}_{a\sim (\pi^1,\pi^2)} \{u^1(a)\},\\
&\mathrm{val}^2[u^2] = \max_{\pi^2\in\Delta(A^2)}\min_{\pi^1\in\Delta(A^1)} \mathbb{E}_{a\sim (\pi^1,\pi^2)} \{u^2(a)\}.
\end{align} 
It is instructive to examine the following thought experiment. Imagine that players are at the edge of the infinite horizon. Then the players' continuation payoff would be determined  by the stage game at state $s$ since there would not be 
any future stages to consider. The unique equilibrium values they would get would be $\val^i[r^i(s,\cdot)]$. Then, at the stage just before the last one, they would have played the strategic-form game $\langle A^1,A^2,Q^1(s,\cdot),Q^2(s,\cdot)\rangle$ at state $s$, where
\begin{align}
Q^i(s,\cdot) = r^i(s,\cdot) + \gamma \sum_{s'\in S} p(s'|s,\cdot) \mathrm{val}^i[r^i(s',\cdot)].
\end{align}
\cite{ref:Shapley53} showed that if we follow this \textit{backward induction}, we can always compute the equilibrium values associated with a stationary equilibrium. To this end, he introduced the operator $\mathcal{T}^i$ defined by
\begin{equation}\label{eq:operator}
(\mathcal{T}^iv^i)(s) := \mathrm{val}^i\left[r^i(s,\cdot) + \gamma \sum_{s'\in S} p(s|s,\cdot) v^i(s')\right],\quad \forall s\in S,
\end{equation}
which is a contraction with respect to the $\ell_{\infty}$-norm when $\gamma \in (0,1)$ since $\mathrm{val}^i$ is a non-expansive mapping, i.e., 
$$
\left|\val^i(u^i) - \val^i(\tilde{u}^i)\right| \leq \max_{a\in A} \left|u^i(a) - \tilde{u}^i(a)\right|,
$$
for any $u^i:A\rightarrow \mathbb{R}$ and $\tilde{u}^i:A\rightarrow \mathbb{R}$,
similar to the maximum function in the Bellman operator. Therefore, the iteration 
\begin{equation}\label{eq:Shapleyupdate}
v_{(n+1)}^i = \mathcal{T}^iv_{(n)}^i,\quad \forall n\geq 0,
\end{equation}
starting from arbitrary $v_{(0)}^i$ converges to the unique fixed point of the operator. Further inspection of the fixed point reveals that it is indeed the equilibrium values of states associated with some stationary equilibrium of the underlying two-player zero-sum stochastic game. There does not exist a counter-part of this iteration for the computation of equilibrium values in general-sum stochastic games, since the value of a game is not uniquely defined for general-sum stochastic games, and involves a fixed point operation, which is hard to compute at each stage of an algorithm. However, Shapley's iteration is still a powerful method to compute equilibrium values in a two-player zero-sum stochastic game. 

In the following section, we examine whether a stationary equilibrium would be realized as a consequence of non-equilibrium adaptation of learning agents as in Section \ref{sec:FP} but now for stochastic games instead of repeated play of the same strategic-form game. 


\section{Learning in Stochastic Games}\label{sec:learning}

Fictitious play dynamics is a best-response type learning dynamics where each player aims to take the best response against the opponent by learning the opponent's strategy based on the history of the play. This stylist learning dynamic can be generalized to stochastic games as players (again erroneously) assume that the opponent plays according to a \textit{stationary} strategy (which depends only on the current state). Hence, they can again form a belief on the opponent's stationary strategy based on the history of the play. Particularly, they can form a belief on the opponent's mixed strategy specific to a state based on the actions taken at that state only due to the stationarity assumption on the opponent's strategy. Given that belief on the opponent's strategy, players can also compute the value of each state-action pair based on \textit{backward induction} since their actions determine both the stage payoff and the continuation payoff by determining the state transitions. Therefore, they essentially play an \textit{auxiliary stage-game} at each stage specific to the current state, which can be represented by $\mathcal{G}_s := \langle A^1,A^2,Q^1(s,\cdot),Q^2(s,\cdot)\rangle$, where the payoff or the \textit{$Q$-function}, $Q^i(s,\cdot):A\rightarrow\mathbb{R}$ is determined according to the backward induction given $\pi^{-i}$ the belief of \player{i} about \player{-i}'s strategy, and therefore, it satisfies the following fixed-point equation
\begin{equation}\label{eq:q}
Q^{i}(s,a) = r^{i}(s,a) + \gamma \sum_{s'\in S} p(s'|s,a)  \max_{a^i\in A^i} \mathbb{E}_{a^{-i}\sim \pi^{-i}(s')}\{Q^i(s',a^1,a^2)\}.
\end{equation}  
For notational convenience, we also define the {\it value function} $v^i:S\rightarrow\mathbb{R}$ by
\begin{equation}\label{eq:v}
v^i(s) := \max_{a^i\in A^i} \mathbb{E}_{a^{-i}\sim \pi^{-i}(s)}\{Q^i(s,a^1,a^2)\}.
\end{equation}


At each stage $k$, player $i$ has a belief on player $-i$'s strategy, which we denote by 
$\hat{\pi}_{k}^{-i}$. Player $i$ also forms a  belief on the payoff function for the auxiliary game, or the $Q$-function, denoted by 
$\hat{Q}_{k}^i$. Let $s$ be the current state of the stochastic game. Then, \player{i} selects her action $a_k^i$ according to
\begin{equation}\label{eq:best}
a_k^i \in \argmax_{a^i\in A^i} \; \mathbb{E}_{a^{-i} \sim \hat{\pi}_k^{-i}(s)}\left\{\hat{Q}_{k}^i(s, a^1,a^2)\right\}.
\end{equation}
Observing the opponent's action $a_k^{-i}$, \player{i} forms her belief on \player{-i}'s strategy for the current state $s$ as a weighted empirical average, which can  be constructed iteratively as 
\begin{equation}\label{eq:piupdate}
\hat{\pi}_{k+1}^{-i}(s) = \hat{\pi}_{k}^{-i}(s) + \alpha_{c_k(s)}(a_k^{-i} - \hat{\pi}_{k}^{-i}(s)).  
\end{equation} 
Here $\alpha_{c} \in [0,1]$ is a step size and it vanishes with $c_k(s)$ indicating the number of visits to state $s$ rather than time. Note that if there was a single state, $c_k(s)$ would correspond to the time, i.e., $c_k(s) = k$, as in the classical fictitious play. The update \eqref{eq:piupdate} can also be viewed as taking a convex combination of the current belief $\hat{\pi}_{k}^2(s)$ and the observed action $a_k^2$ while the step size $ \alpha_{c_k(s)}$ is the (vanishing) weight of the action observed. Vanishing step size as a function of the  number of visits implies that,  the players give less weight to their current belief than the observed action by using a large step size if that state has not been visited many times. This means that the players will still give less weight to their current belief even at later stages if the specific state has not been visited many times, and indicating, they have not been able to strengthen their belief enough to rely more on it.

Simultaneously, \player{i} updates her belief on her own $Q$-function for the current state $s$ according to
\begin{equation}\label{eq:Qupdate}
\hat{Q}_{k+1}^i(s,a) = \hat{Q}_{k}^i(s,a) + \beta_{c_k(s)}\left(r^i(s,a) + \gamma \sum_{s'\in S} p(s'|s,a) \hat{v}_{k}^i(s') - \hat{Q}_{k}^i(s,a)\right),\quad\forall a\in A,
\end{equation}
where we define $\hat{v}^i_k:S\rightarrow \mathbb{R}$ as the value function estimate given by
\begin{equation}\label{eq:vupdate}
\hat{v}_{k}^i(s) = \max_{a^i} \; \mathbb{E}_{a^{-i}_k\sim\hat{\pi}_k^{-i}(s)}\left\{\hat{Q}_{k}^i(s,a^1,a^2)\right\},
\end{equation}
and $\beta_{c}\in[0,1]$ is another step size that also vanishes with $c_k(s)$. Similar to \eqref{eq:piupdate}, the update of the belief on the $Q$-function \eqref{eq:Qupdate} can be viewed as a convex combination of the current belief $\hat{Q}_{k}^i(s,a)$ and the new observation $r^i(s,a) + \gamma \sum_{s'\in S} p(s'|s,a) \hat{v}_{k}^i(s')$. Such vanishing step size again implies that the players are relying on their beliefs more if they have had many chances to strengthen them. 

The \textbf{key feature} of this learning dynamic is that the players update their beliefs on their $Q$-functions at a slower timescale than the update of their beliefs on the opponent strategy. This is consistent with the literature on evolutionary game theory \citep{ref:Ely01,ref:Sandholm01} (which postulates players' choices to be more dynamic than changes in their preferences) since we can view $Q$-functions in auxiliary games as slowly evolving player preferences. Particularly, the two-timescale learning framework implies that the players take smaller and smaller steps at \eqref{eq:Qupdate} than the steps at \eqref{eq:piupdate} such that the ratio of the step sizes, $\beta_c/\alpha_c$, goes to zero with the number of visits to the associated state. Note that this implies that $\beta_c$ goes to zero faster than $\alpha_c$ does, implying slower update of the $Q$-function estimate compared to the opponent's  strategy estimate. This weakens the dependence between evolving beliefs on opponent strategy and $Q$-function.

We say that this {\it two-timescale fictitious play dynamics converge to an equilibrium} if beliefs on opponent strategies converge to a Nash equilibrium which associates with the  auxiliary games while the beliefs on $Q$-functions converge to the $Q$-functions for  a stationary equilibrium of the underlying stochastic game. Particularly, given an equilibrium $\pi_*$, the associated $Q$-function of \player{i} satisfies
$$
Q^i(s,a) = r^i(s,a) + \gamma \sum_{s'\in S} p(s'|s,a) \max_{a^i\in A^i}\mathbb{E}_{a^{-i} \sim \pi_*^{-i}(s)} \{Q^i(s',a^1,a^2)\},\quad\forall (s,a)\in S\times A.
$$ 
Recall that players are playing a dynamically evolving auxiliary game at each state repeatedly, 
but update their beliefs on the $Q$-functions and opponent strategies 
 only when that state is visited. Therefore, the players are updating their beliefs on the opponent strategy and $Q$-function specific to that state only during these visits. 
 Hence, we make the following assumption ensuring that players have sufficient time to revise and improve their beliefs specific to a state.  

\begin{assumption}[Markov Chain]\label{assume:Markov1}
Each state is visited infinitely often.
\end{assumption}

Stochastic games reduce to the repeated play of the same strategic-form game if there exists only one state and the discount factor is zero.  Correspondingly, Assumption \ref{assume:Markov1} always holds in such a case. However, when there are multiple states, Assumption \ref{assume:Markov1} does not necessarily hold, e.g., since some states can be absorbing by preventing transitions to others. In the following, we exemplify four Markov chain configurations with different generality:
\begin{itemize}
\item \textbf{Case $i)$} The probability of transition between any pair of states is positive for \textit{any} action profile. This condition is also known as {\it irreducible stochastic games} \citep{ref:Leslie20}. 
\item \textbf{Case $ii)$} The probability of transition between any pair of states is positive for \textit{at least one} action profile. Case $ii)$ includes Case $i)$ as a special case.\footnote{Another possibility in between Case $i)$ and Case $ii)$ is that the probability of transition between any pair of states is positive for at least one action of one player and any action of the opponent. In other words, the opponent cannot prevent the game to transit from any state to any state.}
\item \textbf{Case $iii)$} There is positive probability that any state can be reached from any state within a finite number of stages for any sequence of action profiles taken during these stages. Case $iii)$ includes Case $i)$ as a special case but not necessarily Case $ii)$.
\item \textbf{Case $iv)$} There is positive probability that any state can be reached from any state within a finite number of stages for at least one sequence of action profiles taken during these stages. 
Case $iv)$ includes Cases $ii)$ and $iii)$ as special cases.\footnote{Another possibility in between Case $iii)$ and Case $iv)$ is that there is positive probability that any state can be reached from any state within a finite number of stages for at least one sequence of actions of one player and for any sequence of actions taken by the opponent during these stages. In other words, the opponent cannot prevent the player to reach any state from any state within a finite number of stages.}
\end{itemize}
Note that Assumption \ref{assume:Markov1} holds under Case $iii)$ but not necessarily under Cases $ii)$ or $iv)$.

Recall that in the classical fictitious play, the beliefs on opponent strategy are formed by  the empirical average of the actions taken by the opponent. The players can also form their  beliefs as a weighted average of the actions while the weights may give more (or less) importance to recent ones depending on the player's preferences, e.g., as in \eqref{eq:piupdate}. In other words, we let $\alpha_c$ take values other than $1/(c+1)$ for $c=0,1,\ldots$. Furthermore, the two-timescale learning scheme imposes that $\beta_c/\alpha_c$ goes to zero as $c$ goes to infinity. In the following, we specify  conditions on step sizes that are sufficient to ensure convergence of the two-timescale fictitious play in two-player zero-sum stochastic games under Assumption \ref{assume:Markov1}. 

\begin{assumption}[Step Sizes]\label{assume:rate1}
The step sizes $\{\alpha_c\}$ and $\{\beta_c\}$ satisfy the following conditions:
\begin{itemize}
\item[(a)] They vanish at a slow enough rate such that
$$
\sum_{c\geq 0} \alpha_c = \sum_{c\geq 0} \beta_c = \infty
$$  
while $\alpha_c\rightarrow 0$ and $\beta_c\rightarrow 0$ as $c\rightarrow \infty$.
\item[(b)] They vanish at two separate timescales such that
$$
\lim_{c\rightarrow \infty} \frac{\beta_c}{\alpha_c} = 0.
$$
\end{itemize}
\end{assumption}

The following theorem shows that the two-timescale fictitious play converges in two-player zero-sum stochastic games under these assumptions.

\begin{theorem}[\cite{ref:Sayin20}]\label{thm:modelbased}
Given a two-player zero-sum stochastic game, suppose that players follow the two-timescale fictitious play dynamics \eqref{eq:piupdate} and \eqref{eq:Qupdate}. Under Assumptions \ref{assume:Markov1} and \ref{assume:rate1}, we have
\begin{equation}
(\hat{\pi}_k^1,\hat{\pi}_k^2) \rightarrow (\pi_*^1,\pi_*^2)\quad\mbox{and}\quad (\hat{Q}_k^1,\hat{Q}_k^2) \rightarrow (Q_*^1,Q_*^2),\quad \mbox{w.p.$1$},
\end{equation}
as $k\rightarrow \infty$ for some stationary equilibrium $\pi_*=(\pi_*^1,\pi_*^2)$ of the underlying stochastic game and $(Q_*^1,Q_*^2)$ denote the associated $Q$-functions.
\end{theorem}

%

Before delving into the technical details of the proof, it is instructive to compare the two-timescale fictitious play with both the classical fictitious play and the Shapley's iteration. For example, the update of $\hat{\pi}^{-i}_k$, described in \eqref{eq:piupdate}, differs from the classical fictitious play dynamics \eqref{equ:FP_1} since the auxiliary game depends on the belief $\hat{Q}^i_k$ while the belief (and therefore the payoffs of the auxiliary games) evolves in time with new observations, quite contrary to the classical scheme \eqref{equ:FP_2}. In general, this constitutes a challenge in directly adopting the convergence analysis for the classical scheme to stochastic games. However, the two-timescale learning scheme weakens this coupling, enabling us to characterize the asymptotic behavior specific to a state separately from the dynamics in other states as if $(\hat{Q}_k^1(s,\cdot),\hat{Q}_k^2(s,\cdot))$ is stationary. 

Moreover, even with the two-timescale learning scheme, we still face a challenge in directly adopting the convergence analysis of fictitious play specific to zero-sum games, e.g., \cite{robinson1951iterative,ref:Harris98}. Particularly, players form beliefs on their $Q$-functions independently based on the backward induction that they will always look for maximizing their utility against the opponent strategy. Due to this independent update, the auxiliary games can deviate from the zero-sum structure even though the underlying game is zero-sum. Hence we do not necessarily have $\hat{Q}_k^1(s,a) + \hat{Q}_k^2(s,a) = 0$ for all $a\in A$ and for each $s\in S$. This poses an important challenge in the analysis since an arbitrary general-sum game does not necessarily have fictitious play property in general. 

Next, we compare the two-timescale fictitious play with Shapley's value iteration. We can list the differences between the update of $\hat{Q}_k^i$, described in \eqref{eq:Qupdate}, and the Shapley's iteration \eqref{eq:Shapleyupdate} as follows:
\begin{itemize}
\item The Shapley's iteration is over the value functions, however, it can be turned into an iteration over the $Q$-functions with the operator
\begin{equation}\label{eq:F}
(\mathcal{F}^iQ^i)(s,a) = r^i(s,a) + \gamma \sum_{s'\in S} p(s'|s,a) \mathrm{val}^i(Q^i(s',\cdot)),\quad \forall (s,a)\in S\times A,
\end{equation}
as derived in \citep{szepesvari1999unified}. The transformed iteration is given by $Q_{(n+1)}^i = \mathcal{F}^iQ_{(n)}^i$ starting from arbitrary $Q_{(0)}^i$. Furthermore, the Shapley's iteration does not involve a step size, however, a step size can be included  if we view $Q_{(n+1)}^i = \mathcal{F}^iQ_{(n)}^i$ as the one 
\begin{equation}
Q_{(n+1)}^i = Q_{(n)}^i + \beta_{(n)} (\mathcal{F}^iQ_{(n)}^i-Q_{(n)}^i)
\end{equation}
with the step size $\beta_{(n)} = 1$ for all $n$.
\item The Shapley's iteration updates the value function at every state at each stage while \eqref{eq:Qupdate} takes place only when the state is visited. Therefore, we face the asynchronous update challenge in the convergence analysis of \eqref{eq:Qupdate} together with \eqref{eq:piupdate}, which can take place only when the associated state is visited. To address this, we can resort to the asynchronous stochastic approximation methods, e.g., see \cite{ref:Tsitsiklis94} (also upcoming Theorem \ref{thm:Tsitsiklis}).
\item More importantly, the convergence of the Shapley's iteration benefits from the contraction property of the operator \eqref{eq:operator} (or its transformed version \eqref{eq:F}) based on the non-expansive mapping $\mathrm{val}^i(\cdot)$. However, in the update \eqref{eq:Qupdate}, we have
$$
\hat{v}^i_k(s) = \max_{a^i\in A^i} \;\mathbb{E}_{a^{-i}\sim \hat{\pi}^{-i}(s)}\{\hat{Q}^i(s,a^1,a^2)\} 
$$
rather than $\mathrm{val}^i(\hat{Q}^i(s,\cdot))$, which need not lead to a contraction.
\end{itemize}

The proof of Theorem \ref{thm:modelbased} follows from exploiting the two-timescale learning scheme to analyze the evolution of the beliefs on opponent strategies specific to a state in isolation as if the beliefs on $Q$-functions are stationary and then showing that $\hat{v}^i_k(s)$ tracks $\mathrm{val}^i(\hat{Q}^i(s,\cdot))$ while addressing the deviation from the zero-sum structure via a novel Lyapunov function construction. The two-timescale learning scheme yields that the limiting flow of the dynamics specific to a state is given by
\begin{align}
&\frac{d\pi^i(s)}{dt} + \pi_s^i \in \argmax_{a^i\in A^i} \mathbb{E}_{a^{-i}\sim \pi^{-i}(s)} \{Q^i(s,a^1,a^2)\}\label{eq:flow}\\
&\frac{dQ^i(s,a)}{dt} = 0,
\end{align} 
for all $(s,a)\in S\times A$ and $i=1,2$. The function \eqref{eq:Lyapunov} presented in \cite{ref:Harris98} for continuous-time best response dynamics in zero-sum games is no longer a valid Lyapunov function since $\sum_{i=1,2}Q^i(s,a)$ is not necessarily zero for all $s$ and $a$. Therefore, we modify this function to characterize the asymptotic behavior of this flow in terms of the deviation from the zero-sum structure, e.g., $\max_{a\in A}|\sum_{i=1,2}Q^i(s,a)|$. The new function is defined by
\begin{align*}
V_*(\pi(s),Q(s,\cdot)) := \left(\sum_{i=1,2} \max_{a^i\in A^i} \mathbb{E}_{a^{-i}\sim \pi^{-i}(s)} \{Q^i(s,a^1,a^2)\} - \lambda \max_{a\in A}\left|\sum_{i=1,2}Q^i(s,a)\right|\right)_+,
\end{align*}
where $\lambda$ is a fixed scalar satisfying $\lambda \in (1,1/\gamma)$. The lower bound on $\lambda$ plays a role in its validity as a Lyapunov function when $\max_{a\in A}\left|\sum_{i=1,2}Q^i(s,a)\right|\neq 0$ while the upper bound will play a role later when we focus on the evolution of $\sum_{i=1,2}Q^i(s,a)$ to show that the sum converges to zero, i.e., the auxiliary stage games become zero-sum, almost surely.

Note that $V_*(\cdot)$ reduces to $V(\cdot)$, described in \eqref{eq:Lyapunov}, if $\sum_{i=1,2}Q^i(s,a)=0$ for all $a\in A$. Furthermore, it is a valid Lyapunov function for any $Q^1(s,\cdot)$ and $Q^2(s,\cdot)$ since we have
\begin{equation} 
\frac{d}{dt}\left(\sum_{i=1,2} \max_{a^i\in A^i} \mathbb{E}_{a^{-i}\sim \pi^{-i}(s)} \{Q^i(s,a)\}\right) = \sum_{i=1,2} Q^i(s,a_*) - \sum_{i=1,2} \max_{a^i\in A^i} \mathbb{E}_{a^{-i}\sim \pi^{-i}(s)} \{Q^i(s,a)\},
\end{equation}
where $a_*=(a_*^1,a_*^2)$ are the maximizing actions in \eqref{eq:flow}, and we always have 
$$
\sum_{i=1,2} Q^i(s,a_*) < \lambda \max_{a\in A}\left|\sum_{i=1,2}Q^i(s,a)\right|
$$
if it is not zero-sum, since $\lambda>1$. In other words, the term inside $(\cdot)_*$ in the new Lyapunov function always decreases along the flow when it is non-negative and cannot be positive once it becomes non-positive.

If we let $\bar{v}_k:=\hat{v}_k^1+\hat{v}_k^2$ and $\bar{Q}_k := \hat{Q}_k^1+\hat{Q}_k^2$, the new Lyapunov function yields that
\begin{equation}\label{eq:upp}
\left(\bar{v}_k(s) - \lambda \max_{a\in A}|\bar{Q}_k(s,a)|\right)_+\rightarrow 0
\end{equation}
as $k\rightarrow \infty$ for each $s\in S$. On the other hand, we always have $\bar{v}_k(s)\geq - \lambda  \max_{a\in A}|\bar{Q}_k(s,a)|$ by the definition of $\hat{v}_k^i$. These bounds imply that $\bar{Q}_k(s,a)\rightarrow 0$, and therefore $\bar{v}_k(s)\rightarrow 0$ for all $(s,a)\in S\times A$, because the evolution of $\bar{Q}_k$ for the current state $s$ is given by
\begin{equation}
\bar{Q}_{k+1}(s,a) = \bar{Q}_{k}(s,a) + \beta_{c_k(s)}\left(\gamma \sum_{s'\in S} p(s'|s,a) \bar{v}_k(s') - \bar{Q}_k(s,a)\right),\quad \forall a\in A
\end{equation}
by \eqref{eq:Qupdate} while the upper bound on $\lambda$ ensures that $\lambda \gamma \in (0,1)$, and therefore, $\bar{Q}_k(s,a)$ contracts at each stage until it converges to zero for all $s\in S$ and $a\in A$. The asynchronous update and the asymptotic upper bound on $\bar{v}_k$, as described in \eqref{eq:upp}, constitute a technical challenge to draw this conclusion, however, they can be addressed via  asynchronous stochastic approximation methods, e.g., see \cite{ref:Tsitsiklis94}. 

Furthermore, the saddle point equilibrium yields that
\begin{align}
\max_{a^1\in A^1}\mathbb{E}_{a^2\sim \hat{\pi}_k^2(s)} \{\hat{Q}_k^1(s,a)\} \geq \mathrm{val}^1(\hat{Q}_k^1(s,\cdot))\geq \min_{a^2\in A^2}\mathbb{E}_{a^1\sim \hat{\pi}_k^1(s)} \{\hat{Q}_k^1(s,a)\}
\end{align}
and the right-hand side is bounded from below by
\begin{align}
\min_{a^2\in A^2}\mathbb{E}_{a^1\sim \hat{\pi}_k^1(s)} \{\hat{Q}_k^1(s,a)\}
&\geq \min_{a^2\in A^2}\mathbb{E}_{a^1\sim \hat{\pi}_k^1(s)} \{-\hat{Q}_k^2(s,a)\} + \min_{a^2\in A^2}\mathbb{E}_{a^1\sim \hat{\pi}_k^1(s)} \{\bar{Q}_k(s,a)\}\\
&\geq -\max_{a^2\in A^2}\mathbb{E}_{a^1\sim \hat{\pi}_k^1(s)} \{\hat{Q}_k^2(s,a)\} - \max_{a\in A}|\bar{Q}_k(s,a)|.
\end{align}
These bounds lead to 
\begin{equation}
0\leq \hat{v}_k^i(s) - \val^i(\hat{Q}_k^i(s,\cdot)) \leq \bar{v}_k(s) + \max_{a\in A}|\bar{Q}_k(s,a)|,
\end{equation}
Since the right-hand side goes to zero as $k\rightarrow \infty$, we have $\hat{v}_k^i(s)$ tracks $\val^i(\hat{Q}_k^i(s,\cdot))$. Based on this tracking result, the update of $\hat{Q}_k^i$ can be viewed as an asynchronous version of the iteration
\begin{equation}
Q_{(n+1)}^i = Q_{(n)}^i + \beta_{(n)}(\mathcal{F}^iQ_{(n)}^i + \epsilon_{(n)}^i - Q_{(n)}^i),
\end{equation}
where the tracking error $\epsilon_{(n)}^i$ is asymptotically negligible almost surely and the operator $\mathcal{F}$, as described in \eqref{eq:F}, is a contraction similar to the Shapley's operator, described in \eqref{eq:operator}. This completes the sketch of the proof for Theorem \ref{thm:modelbased}.

\subsection{Model-Free Learning in Stochastic Games}\label{sec:model_free_SG}

We next consider scenarios where players do not know the transition probabilities and their own stage payoff function, however, they can still observe their stage payoffs (associated with the current action profile), the opponent's action, and the current state visited. Therefore, the players can still form beliefs on opponent strategy and their $Q$-functions. 

The update of the belief on opponent strategy does not depend on the model knowledge. Therefore, the players can update their beliefs $\hat{\pi}_k^{-i}$ as in \eqref{eq:piupdate} also in the model-free case. However, the update of $\hat{Q}_k^i$ necessitates the model knowledge by depending on the stage payoff function and transition probabilities. The same challenge arises also in model-free solution of Markov decision processes (MDPs) - a \textit{single} player version of stochastic games. 

For example, $Q$-learning algorithm, introduced by \cite{ref:Watkins92}, can be viewed as a model-free version of the value iteration in MDPs and the update rule is given by
\begin{equation}\label{eq:Qlearning}
\hat{q}_{k+1}(s,a) = \hat{q}_{k}(s,a) + \beta_k(s,a) \left(r_k + \gamma \max_{\tilde{a}\in A} \hat{q}_{k}(\tilde{s},\tilde{a}) - \hat{q}_k(s,a)\right),
\end{equation}
where the triple $(s,a,\tilde{s})$ denote respectively the current state $s$, current action $a$ and the next state $\tilde{s}$, the payoff $r_k$ corresponds to the payoff received, i.e., $r_k = r(s,a)$, and $\beta_k(s,a)\in[0,1]$ is a step size specific to the state-action pair $(s,a)$. The entries corresponding to the pairs $(s',a')\neq (s,a)$ do not get updated, i.e., $\hat{q}_{k+1}(s',a') = \hat{q}_k(s',a')$.

\cite{ref:Watkins92} provided an ingenious (direct) proof for the almost sure convergence of $Q$-learning algorithm. Alternatively, it is also instructive to establish a connection between $Q$-learning algorithm and the classical value iteration to characterize its convergence properties. For example, the differences between them can be listed as follows:
\begin{itemize}
\item In $Q$-learning, agents use the value function estimate for the next state $\tilde{s}$, i.e., $\hat{v}_k^i(\tilde{s})$, in place of the expected continuation payoff $\sum_{s'\in S} p(s'|s,a) \hat{v}_k^i(s')$. This way, they can sample from the state transition probabilities associated with the current state-action pair by observing the state transitions. Correspondingly, this update takes place only after the environment transitions  to the next state.
\item The update can take place only for the current state-action pair because the agent can sample only from the transition probabilities associated with the current state-action pair by letting the environment do the experimentation.
\end{itemize}
Therefore, the $Q$-learning algorithm can be viewed as an asynchronous $Q$-function version of the value iteration 
\begin{equation}\label{eq:eee}
\hat{q}_{k+1}  = \hat{q}_{k} + \beta_k \left(\mathcal{F}_o\hat{q}_k + \omega_{k+1} - \hat{q}_k\right),
\end{equation}
where the $Q$-function version of the Bellman operator is given by
\begin{equation}
(\mathcal{F}_o\hat{q}_k)(s,a) = r(s,a) + \gamma \sum_{s'\in S} p(s'|s,a) \max_{a'\in A}\hat{q}_{k}(s',a')
\end{equation}
and the stochastic approximation error $\omega_{k+1}$ is defined by
\begin{equation}
\omega_{k+1}(s,a) := \gamma\left(\max_{\tilde{a}\in A}\hat{q}_{k}(\tilde{s},\tilde{a}) - \sum_{s'\in S} p(s'|s,a) \max_{a'\in A}\hat{q}_{k}(s',a')\right)
\end{equation}
with $\tilde{s}$ denoting the next state at stage $k$.
Note that \eqref{eq:eee} turns into an asynchronous update if $\beta_{k}(s,a)$ is just zero when $\hat{q}_k(s,a)$ is not updated. Though these error terms $\{\omega_k\}_{k>0}$ do not form an independent sequence, they form a finite-variance Martingale difference sequence conditioned on the history of parameters. The following well-known result shows that the weighted sum of such Martingale difference sequences vanishes asymptotically almost surely.

\begin{lemma}[\cite{ref:Polyak73}]\label{thm:Polyak}
Let $\{\mathcal{F}_k\}_{k\geq 0}$ be an increasing sequence of $\sigma$-fields. Given a sequence $\{\omega_k\}_{k\geq 0}$, suppose that $\omega_{k-1}$ is $\mathcal{F}_k$-measurable random variable satisfying $\mathbb{E}[\omega_k|\mathcal{F}_k] = 0$ and $\mathbb{E}[\omega^2_k|\mathcal{F}_k]\leq K$ for some $K$. Then, the sequence $\{W_k\}_{k\geq 0}$ evolving according to
\begin{equation}
W_{k+1} = (1-\alpha_k) W_k + \alpha_k \omega_k,
\end{equation}
vanishes to zero asymptotically almost surely, i.e., $\lim_{k\rightarrow\infty} W_k = 0$ with probability $1$, provided that $\alpha_k\in [0,1]$ is a vanishing step size that is $\mathcal{F}_k$-measurable, square-summable $\sum_{k=0}^{\infty}\alpha_k^2 < \infty$ while $\sum_{k=0}^{\infty}\alpha_k = \infty$ with probability $1$.
\end{lemma}

This is a powerful result to characterize the convergence properties of stochastic approximation algorithms having the structure
$$
x_{k+1} = x_k + \alpha_k \left(F(x_k) - x_k+ \omega_k\right)
$$
where $x_k$ is an $n$-dimensional vector, $F:\mathbb{R}^n\rightarrow \mathbb{R}^n$ is a Lipschitz function, $\alpha_k\in[0,1]$ is a step size and $\omega_k$ is a stochastic approximation error term forming a finite-variance Martingale difference sequence conditioned on the history of parameters. Note that every entry of the vector $x_k$ gets updated synchronously. If we also have that the iterate is bounded, we can characterize the convergence properties of this discrete-time update based on its limiting ordinary differential equation via a Lyapunov function formulation \citep{ref:Borkar08}. If the entries do not get updated synchronously, the asynchronous update challenge can be addressed based the \textit{averaging techniques} \citep{ref:Kushner78}. In the case of $Q$-learning, this corresponds to assuming that different state-action pairs occur at well-defined average frequencies, which can be a restriction in practical applications \citep{ref:Tsitsiklis94}. Instead, \citep{ref:Tsitsiklis94} showed that we do not need such an assumption if the mapping $F$ has a contraction-like property based on the asynchronous convergence theory \citep{ref:Bertsekas82,ref:Bertsekas89}.

\begin{theorem}[\cite{ref:Tsitsiklis94}]\label{thm:Tsitsiklis}
Given an MDP, let an agent follow the $Q$-learning algorithm, described in \eqref{eq:Qlearning}, with vanishing step sizes $\beta_k(s,a)\in [0,1]$ satisfying $\sum_{k\geq 0} \beta_k(s,a) = \infty$ and $\sum_{k\geq 0}\beta_k(s,a)^2 < \infty$ for each $(s,a)\in S\times A$. Suppose that the entries corresponding to each $(s,a)$ gets updated infinitely often. Then, we have
\begin{equation}
\hat{q}_k(s,a) \rightarrow q_*(s,a),\quad \mbox{w.p.$1$},
\end{equation}
for each $(s,a)\in S\times A$, as $k\rightarrow\infty$, where $q_*$ is the unique $Q$-function solving the MDP. 
\end{theorem}

\citep{ref:Tsitsiklis94} considered a more general case where agents receive random payoffs. In general, such randomness can result in unbounded parameters. However, this is not the case for $Q$-learning algorithm, i.e., the iterates in the $Q$-learning algorithm remains bounded. Furthermore, the boundedness of the iterates plays a crucial role in the proof of Theorem \ref{thm:Tsitsiklis}. Particularly, consider the deviation between the iterate $\hat{q}_k$ and the unique solution $q_*$, i.e., $\tilde{q}_k = \hat{q}_k - q_*$, which evolves according to
\begin{equation}
\tilde{q}_{k+1} = \tilde{q}_k + \beta_k(\mathcal{F}_o\tilde{q}_k + \omega_{k+1} - \tilde{q}_k)
\end{equation}
by \eqref{eq:eee} and since $\mathcal{F}_oq_* = q_*$. Boundedness of the iterates $\hat{q}_k$ yields that $\tilde{q}_k$ is also bounded. For example, let $|\tilde{q}_k(s,a)|\leq D$ for all $(s,a)$ and $k$. Furthermore, by the contraction property of $\mathcal{F}_o$ with respect to the maximum norm, we have
$$
\max_{(s,a)} |(\mathcal{F}_o\tilde{q}_k)(s,a)| \leq \gamma \max_{(s,a)} |\tilde{q}_k(s,a)|.
$$
Therefore, we can show that the absolute value of new iterates are bounded from above by
\begin{equation}\label{eq:bbb}
|\tilde{q}_{k}(s,a)| \leq Y_k(s,a) + W_{k+1}(s,a),
\end{equation}
where $\{Y_k(s,a)\}_{k\geq 0}$ and $\{W_{k+1}(s,a)\}_{k\geq 0}$ are two sequences evolving, respectively, according to
\begin{equation}
Y_{k+1}(s,a) = (1-\beta_k(s,a))D + \beta_k(s,a) \gamma D 
\end{equation}
starting from $Y_0 = D$, and
\begin{equation}
W_{k+1}(s,a) = (1-\beta_k(s,a))W_k(s,a) + \beta_k(s,a)\omega_{k}(s,a),
\end{equation}
starting from $W_1(s,a) = 0$ for all $(s,a)$. For each $(s,a)$, the sequence $\{Y_k(s,a)\}_{k\geq 0}$ converges to $\gamma D$ while $\{W_{k+1}(s,a)\}_{k\geq 0}$ converges to zero with probability $1$ by Lemma \ref{thm:Polyak} due to the assumptions on the step size and the infinitely often update of every entry. Letting $k\rightarrow\infty$ for both sides of \eqref{eq:bbb}, we obtain that the shifted iterates are asymptotically bounded from above by $\gamma D$. This yields that there exists a stage where the iterates remain bounded from above by $(\gamma +\epsilon)D$ where $\epsilon>0$ is sufficiently small such that $\gamma + \epsilon <1$. By following the same lines, we can find a smaller asymptotic bound on the iterates. Therefore, we can induce that the shifted iterates converge to zero and the iterates converge to the solution of the MDP even with the asynchronous update. 

Similar to the generalization of the value iteration to $Q$-learning for model-free solutions, \cite{ref:Littman94} generalized the Shapley's iteration to \textit{Minimax-$Q$ learning} to compute equilibrium values in two-player zero-sum stochastic games in a model-free way. The update rule is given by
\begin{equation}
\hat{Q}_{k+1}^i(s,a) = \hat{Q}_k^i(s,a) + \beta_k(s,a)\left(r_k^i + \gamma \mathrm{val}^i[\hat{Q}_k^i(\tilde{s})] - \hat{Q}_k^i(s,a)\right),
\end{equation}
for the current state $s$, current action profile $a$, and next state $\tilde{s}$ with a step size $\beta_k(s,a)\in[0,1]$ vanishing sufficiently slow such that $\sum_{k\geq 0}\beta_k(s,a) = \infty$ and $\sum_{k\geq 0}\beta_k(s,a)^2 <\infty$ with probability $1$. The payoff $r_k^i$ corresponds to the payoff received for the current state and action profile, i.e., $r_k^i = r^i(s,a)$. The Minimax-Q algorithm converges to the equilibrium $Q$-functions of the underlying two-player zero-sum stochastic game almost surely if every state and action profile occur infinitely often.

In model-free methods, the assumption that every state-action pair occur infinitely often can be restrictive for practical applications. A remedy to this challenge is that agents explore at random instances by taking any action with uniform probability. Such random exploration results in that every state-action pair gets realized infinitely often if every state is visited infinitely often. Indeed, random exploration will also yield that each state gets visited infinitely often if there is always positive probability that any state is reachable from any state within a finite number of stages for at least one sequence of actions taken during these stages. This corresponds to Case $iv)$ described in Section \ref{sec:learning}. 

In the model-free two-timescale fictitious play, players play the best response in the auxiliary game with probability $(1-\epsilon)$ while experimenting with probability $\epsilon$ by playing any action with uniform probability. They still update the belief on the opponent strategy as in \eqref{eq:piupdate}. Furthermore, they update their beliefs on the $Q$-function for the current state $s$, current action profile $a$ and next state $s'$ triple $(s,a,s')$ according to
\begin{equation}
\hat{Q}_{k+1}^1(s,a) = \hat{Q}_{k}^1(s,a) + \beta_{c_k(s,a)}\left(r^1_k + \gamma \max_{a^1\in A}\mathbb{E}_{a^{2}\sim \hat{\pi}_k^2(s')}\{\hat{Q}_{k}^1(s',a^1,a^2)\} - \hat{Q}_{k}^1(s,a)\right),
\end{equation}
where $\beta_{c_k(s,a)}\in [0,1]$ is a step size vanishing with the number of times $(s,a)$ is realized and the payoff $r_k^1$ corresponds to the payoff associated with the current state $s$ and action profile $a$, i.e., $r_k^1 = r^1(s,a)$. 

Recall that the two-timescale learning scheme plays an important role in the convergence of the dynamics. Particularly, the step size $\alpha_c$ used in the update of the belief $\hat{\pi}_k^{-i}(s)$ goes to zero slower than the step size $\beta_c$ used in the update of the belief $\hat{Q}_k^i(s,\cdot)$. Since both step size depend on the number of visits to the associated state, the assumption that $\beta_c/\alpha_c \rightarrow 0$ as $c\rightarrow \infty$ is sufficient to ensure this timescale separation. However, in the model-free case, the asynchronous update of $\hat{Q}_k^i(s,a)$ for different action profiles can undermine this timescale separation because the step size $\beta_c$ specific to the update of $\hat{Q}_k^i(s,a)$ depends the number of times the state and action profile $(s,a)$, i.e., $c_k(s,a)$, is realized. Therefore, we make the following assumption ensuring that the step size in the update of $\hat{Q}_k^i(s,a)$ vanishes still faster than the step size in the update of $\hat{\pi}_k^{-i}(s)$ as long as $c_k(s,a)$ is comparable with $c_k(s)$, i.e., $\liminf_{k\rightarrow \infty} c_k(s,a)/c_k(s) > 0$ with probability $1$.

\begin{assumption}[Step Sizes]\label{assume:rate2}
The step sizes $\{\alpha_c\}$ and $\{\beta_c\}$ satisfy the following conditions:
\begin{itemize}
\item[(a)] They vanish at a slow enough rate such that
$$
\sum_{c\geq 0} \alpha_c = \sum_{c\geq 0} \beta_c = \infty, \quad \mbox{and}\quad  \sum_{c\geq 0} \alpha_c^2 < \infty,\; \sum_{c\geq 0} \beta_c^2 < \infty
$$  
while $\alpha_c\rightarrow 0$ and $\beta_c\rightarrow 0$ as $c\rightarrow \infty$.\footnote{We have the additional assumption that the step size $\beta_c$ is square summable to ensure that the stochastic approximation error terms have finite variance conditioned on the history of the parameters.}
\item[(b)] The sequence $\{\beta_c\}_{c\geq 0}$ is monotonically decreasing. For any $m \in (0,1]$, we have\footnote{\cite{ref:Perkins12} made a similar assumption that $\sup_c \frac{\beta_{\lfloor mc \rfloor}}{\beta_c}<M$ for all $m\in(0,1)$ and $\frac{\beta_c}{\alpha_c}\rightarrow 0$ for two-timescale asynchronous stochastic approximation.}
$$
\lim_{c\rightarrow \infty}\frac{\beta_{\lfloor mc \rfloor}}{\alpha_c} = 0.
$$
\end{itemize}
\end{assumption}

When we have $\liminf_{k\rightarrow \infty} c_k(s,a)/c_k(s) > 0$ with probability $1$ for all $(s,a)$, the second part of Assumption \ref{assume:rate2} ensures that $\lim_{k\rightarrow \infty} \frac{\beta_{c_k(s,a)}}{\alpha_{c_k(s)}} = 0$ with probability $1$ for all $(s,a)$. Indeed, Assumptions \ref{assume:rate1} and \ref{assume:rate2}  are satisfied for the usual (vanishing) step sizes such as 
$$
\alpha_c = \frac{1}{(c+1)^{\rho_{\alpha}}}\quad\mbox{and}\quad
\beta_c = \frac{1}{(c+1)^{\rho_{\beta}}},
$$
where $1/2< \rho_{\alpha} < \rho_{\beta} \leq 1$.

When players do random experimentation in the model-free case, they do not take the best response with certain probability. Therefore, we do not have convergence to an exact equilibrium as in the model-based case. However, the players still converge to a near equilibrium of the game with linear dependence on the experimentation probability and the following theorem provides an upper bound on this approximation error.

\begin{theorem}[\cite{ref:Sayin20}]
Given a two-player zero-sum stochastic game, suppose that players follow the model-free two-timescale fictitious play dynamics with experimentation probability $\epsilon>0$. Under Assumptions \ref{assume:Markov1} and \ref{assume:rate2}, we have
\begin{align}
&\limsup_{k\rightarrow\infty} |v_k^i(s) - v^i(s)| \leq \epsilon D \frac{1+\gamma}{\gamma (1-\gamma)^2}\\
&\limsup_{k\rightarrow\infty} \max_{a\in A} |\hat{Q}_k^i(s,a) - Q^i(s,a)| \leq \epsilon D \frac{1+\gamma}{(1-\gamma)^2},
\end{align}
with probability $1$, where $D = \frac{1}{1-\gamma}\sum_i \max_{(s,a)} |r^i(s,a)|$, where $v^i_*$ and $Q_*^i$ denote, respectively, the value function and $Q$-function of \player{i} for some stationary equilibrium of the stochastic game.
\end{theorem}

Even though the random experimentation can prevent convergence to an exact equilibrium, it provides an advantage for the applicability of this near-convergence result because every state gets visited infinitely often, and therefore, Assumption \ref{assume:Markov1} holds, if the underlying Markov chain satisfies Case $iv)$, i.e., there is positive probability that any state can be reached from any state within a finite number of stages for at least one sequence of action profiles taken during these stages.

The dynamics can converge to an exact equilibrium also in the model-free case if players let the experimentation probability vanish at certain rate. However, there are technical details that can limit the applicability of the result for Case $iv)$. 

\subsection{Radically Uncoupled Learning in Stochastic Games}

Finally, we consider minimal-information scenarios where players do not even observe the opponent's actions in the model-free case.  Each player can still observe its own stage payoff received and the current state visited. The players also do not know the opponent's action set. Indeed, they may even be oblivious to the presence of an opponent. The learning dynamics under such minimal information case is known as \textit{radically uncoupled learning} in the learning in games literature, e.g., see \citep{ref:Foster06}. 

Without observing the opponent's actions and knowing her action space, players are not able to form beliefs on opponent strategy as in the fictitious play. This challenge is present also in the repeated play of the same strategic-form game. For example, consider the strategic-form game $\langle A^1,A^2,r^1,r^2\rangle$ and define $q^i:A^i\rightarrow\mathbb{R}$ by 
\begin{equation}\label{eq:smallq}
q^i(a^i) := \mathbb{E}_{a^{-i}\sim \pi^{-i}} \{r^i(a^1,a^2)\},\quad\forall a^i\in A^i
\end{equation}
given the opponent's strategy $\pi^{-i}$. Then, the computation of the best response is a simple optimization problem for player $i$, given by
$$
a^i_* \in \argmax_{a^i\in A^i} q^i(a^i).
$$
Player $i$ would be able to compute her best response $a_*^i$ even when she does not know the opponent strategy $\pi^{-i}$ and her payoff function $r^i$ if she knew the function $q^i(\cdot)$. Hence, the question is whether the computation of $q^i(\cdot)$ can be achieved without observing the opponent's action. 


Suppose that players are playing the same strategic-form game repeatedly and player $i$ makes the forward induction that the opponent will play as how he has played in the past similar to the fictitious play dynamics. If that were the case, i.e., the opponent were playing according to a stationary strategy $\pi^{-i}$, then at each stage the payoff received by player $i$ would be the realized payoff $r^i(a^1,a^2)$, where $a^{-i}\sim \pi^{-i}$ and $a^i$ is the current action she has taken. Correspondingly, player $i$ can form a belief about $q^i(a^i)$ for all $a^i\in A^i$ and update $q^i(\cdot)$ associated with the current action based on the payoff she received. For example, let $\hat{q}^i_k$, $a_k^i$ and $r_k^i$ denote, respectively, the belief of player $i$ on $q^i$, her current action and the current payoff she received. Similar to the update of the belief on opponent's strategy, the update of $\hat{q}_k^i$ is given by
$$
\hat{q}^i_{k+1}(a^i) =  \left\{\begin{array}{ll} 
\hat{q}^i_k(a^i) + \alpha_k(a^i) (r_k^i - \hat{q}_k^i(a^i)) & \mbox{if } a^i = a_k^i\\
\hat{q}^i_k(a^i) & \mbox{o.w.} 
\end{array}\right. 
$$
where $\alpha_{k}(a^i)\in[0,1]$ is a vanishing step size specific to the action $a^i$. However, this results in an asynchronous update of $\hat{q}_k$ for different actions quite contrary to the synchronous belief update \eqref{equ:FP_1} in the fictitious play. There is no guarantee that it would converge to an equilibrium even in the zero-sum case. On the other hand, such an asynchrony issue is not present and the update turns out to be synchronous in expectation if players take \textit{smoothed} best response while normalizing the step size by the probability of the current action taken \citep{ref:Leslie05}.   

Given $\hat{q}_k^i$, the smoothed best response $\smooth_k^i\in\Delta(A^i)$ is given by
\begin{equation}\label{eq:smooth}
\smooth_k^i := \argmax_{\mu^i\in \Delta(A^i)} \left(\mathbb{E}_{a^i\sim\mu^i}\{\hat{q}_k^i(a^i)\} + \tau \nu^i(\mu^i)\right),
\end{equation} 
where $\nu^i:\Delta(A^i)\rightarrow\mathbb{R}$ is a smooth and strictly concave function whose gradient is unbounded at the boundary of the simplex $\Delta(A^i)$ \citep{fudenberg1998theory}. The temperature parameter $\tau>0$ controls the amount of perturbation on the smoothed best response. Note that the smooth perturbation ensures that there always exists a unique maximizer in \eqref{eq:smooth}. Since players take smoothed best response rather than best response, we use an equilibrium concept different from the Nash equilibrium. This new definition is known as Nash distribution or quantal response equilibrium \citep{ref:McKelvey95}.

\begin{definition}[Nash Distribution]
We say that a strategy profile $\pi_*$ is a Nash distribution if we have
\begin{align}
\pi_*^i = \argmax_{\mu^i\in \Delta(A^i)} \left(\mathbb{E}_{(a^i,a^{-i})\sim(\mu^i,\pi_*^{-i})}\{r_k^i(a)\} + \tau \nu^i(\mu^i)\right) 
\end{align}
for each $i$.
\end{definition}

An example to the smooth function is $\nu^i(\mu^i) := -\mathbb{E}_{a^i\sim\mu^i}\{\log(\mu^i(a^i))\}$, also known as the entropy \citep{ref:Hofbauer02}, and the associated smoothed best response has the following analytical form:
$$
\smooth_k^i(a^i) = \frac{\exp\left(\hat{q}_k^i(a^i)/\tau\right)}{\sum_{\tilde{a}^i\in A^i} \exp\left(\hat{q}^i_k(\tilde{a}^i)/\tau\right)},
$$ 
which is positive for all $a^i\in A^i$. 

When player $i$ takes her action according to the smoothed best response $\smooth_k^i$, any action will be taken with some positive probability $\smooth_k^i(a^i)>0$. Hence she can update her belief according to 
\begin{equation}\label{eq:indQ}
\hat{q}^i_{k+1}(a^i) =  \left\{\begin{array}{ll} 
\hat{q}^i_k(a^i) + \smooth_k^i(a^i)^{-1}\alpha_k (r_k^i - \hat{q}_k^i(a^i)) & \mbox{if } a^i = a_k^i\\
\hat{q}^i_k(a^i) & \mbox{o.w.} 
\end{array}\right. 
\end{equation}
where $\alpha_k\in(0,1)$ is a step size vanishing with $k$ and not specific to any action. This asynchronous update rule, also known as \textit{individual $Q$-learning}, turns out to be synchronous in the expectation. Particularly, the new update rule is given by
\begin{equation}\label{eq:syncQ}
\hat{q}_{k+1}^i(a^i) = \hat{q}^i_k(a^i) + \alpha_k \left(\mathbb{E}_{a^{-i}\sim\smooth_k^{-i}}\{r^i(a^1,a^2)\} - \hat{q}_k^i(a^i) + \omega_k^i(a^i)\right),\quad\forall a^i\in A^i,
\end{equation}
and $\omega_k^i(a^i)$ is the stochastic approximation error defined by
\begin{align*}
\omega_k^i(a^i) := &\mathbf{1}_{\{a^i=a_k^i\}} \smooth_k^i(a^i)^{-1}(r_k^i - \hat{q}_k^i(a^i)) - \mathbb{E}_{a\sim\smooth_k}\left\{\mathbf{1}_{\{a^i=a_k^i\}} \smooth_k^i(a^i)^{-1}(r_k^i - \hat{q}_k^i(a^i)) \right\},
\end{align*}
where $\smooth_k = (\smooth_k^1,\smooth_k^2)$, because we have 
$$
\mathbb{E}_{a\sim\smooth_k}\left\{\mathbf{1}_{\{a^i=a_k^i\}} \smooth_k^i(a^i)^{-1}(r_k^i - \hat{q}_k^i(a^i)) \right\} = \mathbb{E}_{a^{-i}\sim\smooth_k^{-i}}\{r^i(a^1,a^2)\} - \hat{q}_k^i(a^i).
$$
Furthermore, the stochastic approximation error term forms a Martingale difference sequence conditioned on the history of iterates while the \textit{boundedness} of the iterates ensure that it has finite variance. Therefore, we can invoke Lemma \ref{thm:Polyak} to characterize the convergence properties of \eqref{eq:syncQ} - a rewritten version of \eqref{eq:indQ} with the stochastic approximation term $\omega_k^i$.

\begin{theorem}[\cite{ref:Leslie05}]
In two-player zero-sum (or identical-payoff) strategic-form games played repeatedly, if both player follows the individual $Q$-learning algorithm, described in \eqref{eq:indQ}, then their estimate $\hat{q}_k^i$ converges to $q_*^i$ for all $a^i\in A^i$ satisfying
$$
q^i_*(a^i) = \mathbb{E}_{a^{-i}\sim \pi^{-i}_*} \{r^i(a^1,a^2)\} 
$$
for some Nash distribution $\pi_*=(\pi_*^1,\pi_*^2)$ under the assumption that the iterates remain bounded. Correspondingly, their smoothed best response also converges to $\pi^*$. 
\end{theorem}

Recall that in stochastic games, players are playing an \textit{auxiliary stage-game} specific to the current state $\mathcal{G}_s = \langle A^1,A^2,Q^1(s,\cdot),Q^2(s,\cdot)\rangle$, where $Q^i$ satisfies \eqref{eq:q}. Therefore, in the minimal information case, each player $i$ can form a belief about the associated 
$$
q^i(s,a^i) := \mathbb{E}_{a^{-i}\sim \pi^{-i}(s)} \left\{Q^i(s,a^1,a^2)\right\},
$$ 
which is now specific to state $s$ contrary to \eqref{eq:smallq}, and update it based on the stage payoffs received as in the individual $Q$-learning dynamics. We can view $q^i$ as the local $Q$-function since it is defined over individual actions rather than action profiles. We denote player $i$'s belief on $q^i$ by $\hat{q}_k^i$. Let $s$ be the current state of the stochastic game. Then, player $i$ selects her action $a_k^i$ according to smoothed best response 
$$
\smooth_k^i(s,\cdot) = \argmax_{\mu^i\in\Delta(A^i)} \left(\mathbb{E}_{a^i\sim\mu^i}\{\hat{q}_k^i(s,a^i)\} + \tau \nu^i(\mu^i)\right),
$$
i.e., $a_k^i\sim \smooth_k^i(s,\cdot)$. The smoothed best response depends only on the belief on the local $Q$-function, i.e., $\hat{q}_k^i(s,\cdot)$. Observing the stage reward $r_k^i$ and the next state $s'$, player $i$ can update her belief according to 
\begin{equation}\label{eq:indQ2}
\hat{q}^i_{k+1}(s,a^i) =  \left\{\begin{array}{ll} 
\hat{q}^i_k(s,a^i) + \smooth_k^i(s,a^i)^{-1}\alpha_{c_k(s)} (r_k^i + \gamma \hat{v}_k^i(s') - \hat{q}_k^i(s,a^i)) & \mbox{if } a^i = a_k^i\\
\hat{q}^i_k(s,a^i) & \mbox{o.w.} 
\end{array}\right. 
\end{equation}
where $\alpha_c\in(0,1)$ is a vanishing step size and recall that $c_k(s)$ denotes the number of visits to state $s$ until and including stage $k$. The update \eqref{eq:indQ2} differs from \eqref{eq:indQ} due to the additional term $\gamma \hat{v}_k^i(s')$ corresponding to an unbiased estimate of the continuation payoff in the model-free case. Due to this additional term, the individual $Q$-learning dynamics in auxiliary stage-games specific to each state are coupled with each other. A two-timescale learning framework can weaken this coupling if players estimate $\hat{v}_k^i$ at a slower timescale according to
\begin{equation}\label{eq:indv}
\hat{v}_{k+1}^i(s) = \hat{v}_k^i(s) + \beta_{c_k(s)} \left(\mathbb{E}_{a^i\sim\smooth_k^i(s,\cdot)}\{\hat{q}_k^i(s,a^i)\} - \hat{v}_k^i(s)\right),
\end{equation}
where $\beta_c\in(0,1)$ is a vanishing step size that goes to zero faster than $\alpha_c$, rather than $\hat{v}_k^i(s) = \mathbb{E}_{a^i\sim\smooth_k^i(s,\cdot)}\{\hat{q}_k^i(s,a^i)\}$.

This decentralized $Q$-learning dynamics, described in \eqref{eq:indQ2} and \eqref{eq:indv}, have convergence properties similar to the two-timescale fictitious play even in this minimal information case. Furthermore, random exploration is inherent in the smoothed best response. Therefore, Assumption \ref{assume:Markov1} holds  if the underlying Markov chain satisfies Case $iv)$. However, due to the smoothed best response, the dynamics does not necessarily converge to an exact Nash equilibrium.

\begin{theorem}[\cite{ref:Sayin21}]\label{thm:minimal}
Given a two-player zero-sum stochastic game, suppose that players follow the decentralized $Q$-learning dynamics. In addition to Assumptions \ref{assume:Markov1} and \ref{assume:rate2}, we assume that $\sum_{c\geq 0}\alpha_c^2<\infty$ and the iterates are bounded. Let $Q_*^i$ and $v_*^i$ denote the unique equilibrium $Q$-function and value function of player $i$. Then, we have 
\begin{align}
\limsup_{k\rightarrow\infty}|\hat{v}_{k}^i(s) - v_{*}^i(s)| \leq \tau \log(|A^1||A^2|)g(\gamma),\label{equ:result_to_eps}
\end{align}
for all $(i,s)\in\{1,2\}\times S$, with probability $1$, where $g(\lambda) := \frac{2+\lambda-\lambda\gamma}{(1-\lambda\gamma)(1-\gamma)}$ with some $\lambda\in(1,1/\gamma)$.

Furthermore, let $\hat{\pi}_{k}^i(s)\in \Delta(A^i)$ be the weighted time-average of the smoothed best response updated as
$$
\hat{\pi}_{k+1}^i(s)=\hat{\pi}_{k}^i(s) + \mathbf{1}_{\{s=s_k\}}\alpha_{c_k(s)}\left(\smooth_k^i(s,\cdot) - \hat{\pi}_{k}^i(s)\right).
$$
Then, we have 
\begin{align}
\limsup_{k\rightarrow\infty}\Big|\max_{a^i\in A^i}\mathbb{E}_{a^{-i}\sim\hat{\pi}_{k}^{-i}(s)}\{Q_{*}^i(s,a)\} - v_{\pi_*}^i(s)\Big| \leq \tau \log(|A^1||A^2|)h(\gamma),
\end{align}
for all $(i,s)\in\{1,2\}\times S$, w.p. $1$, where $h(\gamma) := g(\gamma)(1+\gamma)-1$. In other words, these weighted-average strategies  converge to near Nash equilibrium strategies of the stochastic game. 
\end{theorem}

The iterates would be bounded inherently if players update the local $Q$-function \eqref{eq:indQ} by thresholding the step $\smooth_k^i(a^i)^{-1}\alpha_{c_k(s)}$ from above by $1$. Furthermore, the dynamics could converge to an exact equilibrium if players let their temperature parameter $\tau>0$ vanishes over time at a certain rate, e.g., see \citep{ref:Sayin21}. With vanishing temperature, Assumption \ref{assume:Markov1} holds if the underlying Markov chain satisfies Case $iii)$. 
 
 
\section{Other Learning Algorithms}\label{sec:other}
 
Previous sections have focused on a  detailed description of best-response/fictitious-play type learning dynamics, together with $Q$-learning dynamics, for stochastic games. In this section, we summarize several other algorithms in the learning in games literature, with a focus on independent/decentralized learning for stochastic games (also belonging to the area of {\it multi-agent reinforcement learning} in the machine learning literature). 
 
\subsection{Classical Algorithms} 
 
For stochastic games, other than $Q$-learning-type algorithms presented  in \S\ref{sec:model_free_SG}, \cite{borkar2002reinforcement} also established the asymptotic convergence of an actor-critic algorithm to a weaker notion of generalized Nash equilibrium. Another early work \cite{brafman2002r} proposed {\tt R-MAX}, an  optimism-based RL algorithm for average-reward two-player zero-sum stochastic games, with polynomial time convergence guarantees. However, convergence to the actual Nash equilibrium is not guaranteed from the regret definition in the paper.  

For strategic-form  games, besides fictitious play, several  other  \emph{decentralized} learning dynamics have also been thoroughly studied. A particular example is the \emph{no-regret} learning algorithms\footnote{See \cite{cesa2006prediction}  for formal definitions and results of no-regret learning.} from the online learning literature. 
It is a folklore theorem that:  If both players of a game use some no-regret learning dynamics to adapt their strategies to their opponent's strategies, then the time-average strategies of the players  constitute a Nash equilibrium of the zero-sum strategic-form game 
\citep{cesa2006prediction,roughgarden2010algorithmic}. 
Popular no-regret   dynamics include multiplicative weights update \citep{littlestone1994weighted,freund1999adaptive}, online gradient descent \citep{zinkevich2003online}, and their generalizations  \citep{shalev2011online,mcmahan2011follow}.   These no-regret learning dynamics are \emph{uncoupled} in that  a player's dynamics does  not explicitly rely on the payoffs of other players \citep{hart2003uncoupled}. They are also posited to be a rational model of players' rational behavior \citep{roughgarden2009intrinsic,syrgkanis2013composable}. 
In addition, \cite{ref:Leslie05}  proposed individual $Q$-learning, a fully decentralized  learning dynamics where each player's update rule  requires no observation of the opponent's actions, with convergence to the Nash equilibrium \emph{distribution} of certain two-player games. Notably, these decentralized learning dynamics are only known to be effective for strategic-form games.	

\subsection{Multi-Agent Reinforcement Learning}\label{sec:MARL}
  
There has been a flurry of recent works on multi-agent RL in stochastic games with focuses on \emph{non-asymptotic} performance guarantees.  \cite{perolat2015approximate,pmlr-v54-perolat17a} proposed batch RL algorithms to find an approximate Nash equilibrium using approximate dynamic programming analysis. \cite{wei2017online} studied  \emph{online}  RL,  where only one of the player is controlled,  and develops the {\tt UCSG} algorithm with sublinear regret guarantees  that improves the results in  \cite{brafman2002r}, though still without guarantees of finding the Nash equilibrium. Subsequently, \cite{sidford2019solving} provided near-optimal sample complexity for solving \emph{turn-based} two-player zero-sum finite stochastic games, when a generative model that enables sampling from any state-action pair is available. Under the same setting, the near-optimal sample complexity for general two-player zero-sum finite stochastic games was then established in \cite{zhang2020model}. Without a generative model, \cite{bai2020provable,xie2020learning} presented optimistic value iteration-based RL algorithms for two-player zero-sum stochastic games, with efficient exploration of the environment, and finite-time regret guarantees. The two players need some coordination to perform the algorithms, and the focus in these two works is the \emph{finite-horizon episodic} setting. Later, \cite{bai2020near} and \cite{liu2020sharp} provided tighter regret bounds for the same setting, with model-free and model-based RL methods, respectively. \cite{liu2020sharp} has also studied the general-sum setting, with finite-sample guarantees for finding the Nash equilibrium, assuming some computation oracle for finding the equilibrium of general-sum strategic-form games at each iteration. Contemporaneously, \cite{jin2021power,huang2021towards}  studied multi-agent RL with \emph{function approximation} in finite-horizon episodic zero-sum stochastic games, with also the optimism principle and regret guarantees. 

In addition, \emph{policy-based} RL algorithms  have also been  developed for solving stochastic games. \cite{zhang2019policyb,bu2019global} developed double-loop policy gradient methods for solving zero-sum linear quadratic dynamic games, a special case of zero-sum stochastic games with linear transition dynamics and quadratic cost functions, with convergence guarantees to the Nash equilibrium. Later, \cite{zhao2021provably} also studied double-loop policy gradient methods for zero-sum stochastic games with general function approximation. Note that these double-loop algorithms are not symmetric in that they  require one of the players to wait the opponent to update her policy parameter multiple steps while updating her own policy for one step,  which necessarily requires some coordination between players. Finally, \cite{shah2020reinforcement} developed an Explore-Improve-Supervise approach, which combines ideas from Monte-Carlo Tree Search and  Nearest Neighbors methods, to find the approximate Nash equilibrium value of \emph{continuous-space} turn-based  zero-sum stochastic games. The two players are coordinated to learn the minimax value jointly. 

Notably, as minimax $Q$-learning, these multi-agent RL algorithms are mostly focused on the \emph{computational} aspect of learning in stochastic games: compute the Nash equilibrium without knowing the model, using possibly as few samples as possible. Certain level of coordination among the players is either explicitly or implicitly assumed when implementing these algorithms, even for the zero-sum  setting where the players compete against each other. For human-like self-interested players, these update   rules may not be sufficiently rational and natural to execute. Indeed, as per \cite{bowling2001rational}, a preferable multi-agent RL algorithm should be both \emph{rational} and \emph{convergent}: a rational algorithm ensures that the iterates converge to the opponent's best-response if the opponent converges to a stationary policy; while a convergent algorithm ensures convergence to some equilibrium if all the agents apply the learning dynamics. In general, a rational algorithm, in which each player   \emph{adapts} to the (possibly non-stationary) behavior of other players and uses only \emph{local}  information she observes without the aid of any central coordinator, does not lead to the equilibrium of the game. In fact, investigating whether a game-theoretical equilibrium can be realized  as a result of non-equilibrium adaptation dynamics is the core topic in the literature of \emph{learning in games} \citep{fudenberg1998theory}. These multi-agent RL works have thus motivated our study of independent learning dynamics presented in \S\ref{sec:learning}. 

\subsection{Decentralized Learning in Stochastic Games}

Decentralized learning in stochastic games has attracted increasing research interest lately. In \cite{arslan2017decentralized}, decentralized $Q$-learning has been proposed  for \emph{weakly acyclic} stochastic games, which include stochastic teams (identical-interest stochastic games) as a special case. The update rule for each player does not need to observe the opponent players' actions, and is even oblivious to the presence of other players. However, the players are implicitly coordinated to explore every multiple iterations (in the exploration phase) without changing their policies, in order to create a stationary environment for each player. The key feature of the update rule is to restrict player strategies to stationary pure strategies. Since there are only finitely many stationary pure strategy, players can create a huge-game matrix for each stationary pure strategy and a pure-strategy equilibrium always exists when this huge-game is weakly acyclic with respect to best response. However, in the model-free case, players do not know the payoffs of this huge-game and the two-phase update rule addresses this challenge.
\cite{perolat2018actor} developed actor-critic type learning dynamics that are decentralized and of fictitious-play type, where the value functions are estimated at a faster timescale (in the critic step), and the policy is improved at a slower one (in the actor step). Nonetheless, the learning dynamics only applies to a special class of stochastic games with a ``multistage'' structure, in which each state can only be visited once.  In \cite{daskalakis2020independent}, an independent policy gradient method was investigated for zero-sum stochastic games with convergence rate analysis, where two players use \emph{asymmetric} stepsizes in their updates with one updates faster than the other. This implicitly requires some coordination between players to determine who shall update faster. Contemporaneously, \cite{tian2020provably} studied \emph{online}  RL in unknown stochastic games, where only one player is controlled and the update rule is fully decentralized. The work focused on the efficient \emph{exploration} aspect of multi-agent RL, by establishing the regret\footnote{The regret defined in \cite{tian2020provably} is weaker than the normal one with the \emph{best-in-hindsight} comparator. See \cite[Sec. 2]{tian2020provably} for a  detailed comparison. } guarantees of the proposed update rule. The work considered only the finite-horizon episodic setting, and it is also unclear if the learning dynamics converge to any equilibrium when all players apply it.\footnote{The same update rule with different stepsize and bonus choices and a certified policy technique, however, can return a non-Markovian approximate Nash equilibrium policy pair in the zero-sum setting; see \cite{bai2020near}, and the very recent and more complete treatment \cite{jin2021v},  for more details.}

With symmetric and decentralized learning dynamics, \cite{ref:Leslie20,wei2021last,cen2021fast} are to the best our knowledge the latest efforts on learning in stochastic games. \cite{ref:Leslie20} studied  \emph{continuous-time} best-response dynamics for zero-sum stochastic games, with a \emph{two-timescale} update rule: at the slower timescale, a single continuation payoff (common among the players) is updated, representing time average of auxiliary game payoffs up to time $k$; at the faster timescale, each player updates its strategy in the direction of its best response to opponent's current strategy in the auxiliary game.  The common continuation payoff update ensures that the auxiliary game is always zero-sum, allowing the use of the techniques for the strategic-form  game setting \citep{ref:Harris98}. The  dynamics update the  mixed strategies at every state at every time.  Alternatively, the work also considered  a continuous-time embedding of the \emph{actual play} of the stochastic game where game transitions according to a controlled continuous-time Markov chain. Both  \cite{wei2021last} and \cite{cen2021fast} studied the genuine  infinite-horizon discounted zero-sum stochastic games, and provided \emph{last-iterate} convergence rate  guarantees to approximate Nash equilibrium. To this end, \cite{wei2021last} developed an optimistic variant of gradient descent-ascent update rule; while \cite{cen2021fast} focused on the \emph{entropy-regularized} stochastic games, and advocated the use of policy extragradient methods. Though theoretically strong and appealing, these update rules assume either exact access or sufficiently accurate estimates of the continuation payoffs under instantaneous joint strategies and/or the instantaneous strategy of the opponent. In particular, to obtain finite-time bounds, the players are coordinated to interact multiple steps to estimate the continuation payoffs in the learning setting \citep{wei2021last}.  

By and large, ever since the introduction of  fictitious play \citep{brown1951iterative} and stochastic games  \citep{ref:Shapley53}, it remains a long-standing problem whether an equilibrium in a stochastic game can be realized as an outcome of some natural and decentralized non-equilibrium adaptation, e.g., fictitious play (except the contemporaneous work \cite{ref:Leslie20} with some continuous-time embeddings). Hence, our solutions in \S\ref{sec:learning} serve as an  initial attempt towards settling  the  argument positively. 


\section{Conclusions and Open Problems}\label{sec:conclusion}

In this review paper, we introduced multi-agent dynamic learning in stochastic games, an increasingly active research area where artificial intelligence, specifically reinforcement learning, meets game theory. We have presented the fundamentals and background of the problem, followed by our recent advances in this direction{, with a focus on studying {\it independent learning} dynamics}.  We believe our work has opened up fruitful directions for future research, on developing more natural and rational multi-agent learning dynamics for stochastic games. In particular, several future/ongoing research directions include: 1) establishing convergence guarantees of our independent  learning dynamics for other stochastic games, e.g., identical-interest ones; 2) establishing non-asymptotic convergence guarantees of our learning dynamics, or other independent  learning dynamics, for stochastic games; 3) developing natural learning dynamics that also account for the large state-action spaces in practical stochastic games, e.g., via function approximation techniques. 

\end{spacing}

\newpage
\small
\begin{spacing}{1}
\bibliographystyle{plainnat}
\bibliography{myref,MARL_Springer_1,MARL_Springer_2,RL,mybibfile,mybibfile_2}

\begin{thebibliography}{91}
\providecommand{\natexlab}[1]{#1}
\providecommand{\url}[1]{\texttt{#1}}
\expandafter\ifx\csname urlstyle\endcsname\relax
  \providecommand{\doi}[1]{doi: #1}\else
  \providecommand{\doi}{doi: \begingroup \urlstyle{rm}\Url}\fi

\bibitem[Arslan and Y{\"u}ksel(2017)]{arslan2017decentralized}
G.~Arslan and S.~Y{\"u}ksel.
\newblock Decentralized \relax{Q}-learning for stochastic teams and games.
\newblock \emph{IEEE Transactions on Automatic Control}, 62\penalty0
  (4):\penalty0 1545--1558, 2017.

\bibitem[Ba\c{s}ar and Olsder(1999)]{ref:Basar99}
T.~Ba\c{s}ar and G.~J. Olsder.
\newblock \emph{Dynamic Noncooperative Game Theory}.
\newblock Classics in Applied Mathematics. SIAM, 2nd edition, 1999.

\bibitem[Bai and Jin(2020)]{bai2020provable}
Y.~Bai and C.~Jin.
\newblock Provable self-play algorithms for competitive reinforcement learning.
\newblock In \emph{International Conference on Machine Learning}, pages
  551--560. PMLR, 2020.

\bibitem[Bai et~al.(2020)Bai, Jin, and Yu]{bai2020near}
Y.~Bai, C.~Jin, and T.~Yu.
\newblock Near-optimal reinforcement learning with self-play.
\newblock In \emph{Advances in Neural Information Processing Systems},
  volume~33, 2020.

\bibitem[Benaim et~al.(2005)Benaim, Hofbauer, and Sorin]{ref:Benaim05}
M.~Benaim, J.~Hofbauer, and S.~Sorin.
\newblock Stochastic approximations and differential inclusions.
\newblock \emph{SIAM Journal on Control and Optimization}, 44\penalty0
  (1):\penalty0 328--348, 2005.

\bibitem[Berger(2005)]{ref:Berger05}
U.~Berger.
\newblock Fictitious play in 2xn games.
\newblock \emph{Journal of Economic Theory}, 120\penalty0 (2):\penalty0
  139--154, 2005.

\bibitem[Berger(2008)]{ref:Berger08}
U.~Berger.
\newblock Learning in games with strategic complementarities revisited.
\newblock \emph{Journal of Economic Theory}, 143\penalty0 (1):\penalty0
  292--301, 2008.

\bibitem[Bertsekas(1982)]{ref:Bertsekas82}
D.~P. Bertsekas.
\newblock Distributed dynamic programming.
\newblock \emph{IEEE Transactions on Automatic Control}, AC-27:\penalty0
  610--616, 1982.

\bibitem[Bertsekas and Tsitsiklis(1989)]{ref:Bertsekas89}
D.~P. Bertsekas and J.~N. Tsitsiklis.
\newblock \emph{Parallel and Distributed Computation: Numerical Methods}.
\newblock Prentice Hall New Jersey, 1989.

\bibitem[Borkar(2002)]{borkar2002reinforcement}
V.~S. Borkar.
\newblock Reinforcement learning in {M}arkovian evolutionary games.
\newblock \emph{Advances in Complex Systems}, 5\penalty0 (01):\penalty0 55--72,
  2002.

\bibitem[Borkar(2008)]{ref:Borkar08}
V.~S. Borkar.
\newblock \emph{Stochastic Approximation: A Dynamical Systems Viewpoint}.
\newblock Hindustan Book Agency, 2008.

\bibitem[Bowling and Veloso(2001)]{bowling2001rational}
M.~Bowling and M.~Veloso.
\newblock Rational and convergent learning in stochastic games.
\newblock In \emph{International Joint Conference on Artificial Intelligence},
  volume~17, pages 1021--1026, 2001.

\bibitem[Brafman and Tennenholtz(2002)]{brafman2002r}
R.~I. Brafman and M.~Tennenholtz.
\newblock R-max-{A} general polynomial time algorithm for near-optimal
  reinforcement learning.
\newblock \emph{Journal of Machine Learning Research}, 3\penalty0
  (Oct):\penalty0 213--231, 2002.

\bibitem[Brown(1951)]{brown1951iterative}
G.~W. Brown.
\newblock Iterative solution of games by fictitious play.
\newblock \emph{Activity Analysis of Production and Allocation}, 13\penalty0
  (1):\penalty0 374--376, 1951.

\bibitem[Bu et~al.(2019)Bu, Ratliff, and Mesbahi]{bu2019global}
J.~Bu, L.~J. Ratliff, and M.~Mesbahi.
\newblock Global convergence of policy gradient for sequential zero-sum linear
  quadratic dynamic games.
\newblock \emph{arXiv preprint arXiv:1911.04672}, 2019.

\bibitem[Busoniu et~al.(2008)Busoniu, Babuska, and
  Schutter]{busoniu2008comprehensive}
L.~Busoniu, R.~Babuska, and B.~De Schutter.
\newblock A comprehensive survey of multi-agent reinforcement learning.
\newblock \emph{IEEE Transactions on Systems, Man, And Cybernetics-Part C:
  Applications and Reviews}, 38\penalty0 (2):\penalty0 156--172, 2008.

\bibitem[Cen et~al.(2021)Cen, Wei, and Chi]{cen2021fast}
S.~Cen, Y.~Wei, and Y.~Chi.
\newblock Fast policy extragradient methods for competitive games with entropy
  regularization.
\newblock \emph{arXiv preprint arXiv:2105.15186}, 2021.

\bibitem[Cesa-Bianchi and Lugosi(2006)]{cesa2006prediction}
N.~Cesa-Bianchi and G.~Lugosi.
\newblock \emph{Prediction, Learning, and Games}.
\newblock Cambridge University Press, 2006.

\bibitem[Claus and Boutilier(1998)]{claus1998dynamics}
C.~Claus and C.~Boutilier.
\newblock The dynamics of reinforcement learning in cooperative multiagent
  systems.
\newblock In \emph{Conference on Artificial Intelligence}, pages 746--752,
  1998.

\bibitem[Condon(1990)]{condon1990algorithms}
A.~Condon.
\newblock On algorithms for simple stochastic games.
\newblock \emph{Advances in Computational Complexity Theory}, 13:\penalty0
  51--72, 1990.

\bibitem[Daskalakis et~al.(2020)Daskalakis, Foster, and
  Golowich]{daskalakis2020independent}
C.~Daskalakis, D.~J. Foster, and N.~Golowich.
\newblock Independent policy gradient methods for competitive reinforcement
  learning.
\newblock In \emph{Advances in Neural Information Processing Systems}, 2020.

\bibitem[Ely and Yilankaya(2001)]{ref:Ely01}
J.~C. Ely and O.~Yilankaya.
\newblock Nash equilibrium and the evolution of preferences.
\newblock \emph{Journal of Economic Theory}, 97\penalty0 (2):\penalty0
  255--272, 2001.

\bibitem[Filar and Vrieze(2012)]{filar2012competitive}
J.~Filar and K.~Vrieze.
\newblock \emph{Competitive {M}arkov {D}ecision {P}rocesses}.
\newblock Springer Science \& Business Media, 2012.

\bibitem[Fink(1964)]{ref:Fink64}
A.~M. Fink.
\newblock Equilibrium in stochastic n-person game.
\newblock \emph{Journal of Science Hiroshima University Series A-I},
  28:\penalty0 89--93, 1964.

\bibitem[Foster and Young(2006)]{ref:Foster06}
D.~P. Foster and H.~P. Young.
\newblock Regret testing: learning to play {N}ash equilibrium without knowing
  you have an opponent.
\newblock \emph{Theoretical Economics}, 1:\penalty0 341--367, 2006.

\bibitem[Freund and Schapire(1999)]{freund1999adaptive}
Y.~Freund and R.~E. Schapire.
\newblock Adaptive game playing using multiplicative weights.
\newblock \emph{Games and Economic Behavior}, 29\penalty0 (1-2):\penalty0
  79--103, 1999.

\bibitem[Fudenberg and Kreps(1993)]{fudenberg1993learning}
D.~Fudenberg and D.~M. Kreps.
\newblock Learning mixed equilibria.
\newblock \emph{Games and Economic Behavior}, 5\penalty0 (3):\penalty0
  320--367, 1993.

\bibitem[Fudenberg and Levine(1995)]{fudenberg1995consistency}
D.~Fudenberg and D.~K. Levine.
\newblock Consistency and cautious fictitious play.
\newblock \emph{Journal of Economic Dynamics and Control}, 19\penalty0
  (5-7):\penalty0 1065--1089, 1995.

\bibitem[Fudenberg and Levine(1998)]{fudenberg1998theory}
D.~Fudenberg and D.~K. Levine.
\newblock \emph{The Theory of Learning in Games}, volume~2.
\newblock MIT press, 1998.

\bibitem[Gu et~al.(2017)Gu, Holly, Lillicrap, and Levine]{gu2017deep}
S.~Gu, E.~Holly, T.~Lillicrap, and S.~Levine.
\newblock Deep reinforcement learning for robotic manipulation with
  asynchronous off-policy updates.
\newblock In \emph{IEEE International Conference on Robotics and Automation},
  pages 3389--3396. IEEE, 2017.

\bibitem[Harris(1998)]{ref:Harris98}
C.~Harris.
\newblock On the rate of convergence of continuous-time fictitious play.
\newblock \emph{Games and Economic Behavior}, 22:\penalty0 238--259, 1998.

\bibitem[Hart and Mas-Colell(2003)]{hart2003uncoupled}
S.~Hart and A.~Mas-Colell.
\newblock Uncoupled dynamics do not lead to {N}ash equilibrium.
\newblock \emph{American Economic Review}, 93\penalty0 (5):\penalty0
  1830--1836, 2003.

\bibitem[Hofbauer and Sandholm(2002{\natexlab{a}})]{hofbauer2002global}
J.~Hofbauer and W.~H. Sandholm.
\newblock On the global convergence of stochastic fictitious play.
\newblock \emph{Econometrica}, 70\penalty0 (6):\penalty0 2265--2294,
  2002{\natexlab{a}}.

\bibitem[Hofbauer and Sandholm(2002{\natexlab{b}})]{ref:Hofbauer02}
J.~Hofbauer and W.~H. Sandholm.
\newblock On the global convergence of stochastic fictitious play.
\newblock \emph{Econometrica}, 70:\penalty0 2265--2294, 2002{\natexlab{b}}.

\bibitem[Huang et~al.(2021)Huang, Lee, Wang, and Yang]{huang2021towards}
B.~Huang, J.~D. Lee, Z.~Wang, and Z.~Yang.
\newblock Towards general function approximation in zero-sum markov games.
\newblock \emph{arXiv preprint arXiv:2107.14702}, 2021.

\bibitem[Jin et~al.(2021{\natexlab{a}})Jin, Liu, Wang, and Yu]{jin2021v}
C.~Jin, Q.~Liu, Y.~Wang, and T.~Yu.
\newblock V-learning--{A} simple, efficient, decentralized algorithm for
  multiagent {RL}.
\newblock \emph{arXiv preprint arXiv:2110.14555}, 2021{\natexlab{a}}.

\bibitem[Jin et~al.(2021{\natexlab{b}})Jin, Liu, and Yu]{jin2021power}
C.~Jin, Q.~Liu, and T.~Yu.
\newblock The power of exploiter: {P}rovable multi-agent {RL} in large state
  spaces.
\newblock \emph{arXiv preprint arXiv:2106.03352}, 2021{\natexlab{b}}.

\bibitem[Kushner and Clark(1978)]{ref:Kushner78}
H.~J. Kushner and D.~S. Clark.
\newblock \emph{Stochastic Approximation Methods for Constrained and
  Unconstrained Systems}.
\newblock Springer Verlag, 1978.

\bibitem[Leslie and Collins(2005)]{ref:Leslie05}
D.~S. Leslie and E.~J. Collins.
\newblock Individual {Q}-learning in normal form games.
\newblock \emph{SIAM J. Control Optim.}, 44\penalty0 (2):\penalty0 495--514,
  2005.

\bibitem[Leslie et~al.(2020)Leslie, Perkins, and Xu]{ref:Leslie20}
D.~S. Leslie, S.~Perkins, and Z.~Xu.
\newblock Best-response dynamics in zero-sum stochastic games.
\newblock \emph{Journal of Economic Theory}, 189, 2020.

\bibitem[Littlestone and Warmuth(1994)]{littlestone1994weighted}
N.~Littlestone and M.~K. Warmuth.
\newblock The weighted majority algorithm.
\newblock \emph{Information and Computation}, 108\penalty0 (2):\penalty0
  212--261, 1994.

\bibitem[Littman(1994)]{ref:Littman94}
M.~L. Littman.
\newblock Markov games as a framework for multi-agent reinforcement learning.
\newblock In \emph{International Conference on Machine Learning}, pages
  157--163, 1994.

\bibitem[Liu et~al.(2021)Liu, Yu, Bai, and Jin]{liu2020sharp}
Q.~Liu, T.~Yu, Y.~Bai, and C.~Jin.
\newblock A sharp analysis of model-based reinforcement learning with
  self-play.
\newblock In \emph{International Conference on Machine Learning}, pages
  7001--7010, 2021.

\bibitem[Maskin and Tirole(1988{\natexlab{a}})]{maskin1988theory}
E.~Maskin and J.~Tirole.
\newblock A theory of dynamic oligopoly, {I}: {O}verview and quantity
  competition with large fixed costs.
\newblock \emph{Econometrica: Journal of the Econometric Society}, pages
  549--569, 1988{\natexlab{a}}.

\bibitem[Maskin and Tirole(1988{\natexlab{b}})]{maskin1988theoryb}
E.~Maskin and J.~Tirole.
\newblock A theory of dynamic oligopoly, {II: P}rice competition, kinked demand
  curves, and edgeworth cycles.
\newblock \emph{Econometrica: Journal of the Econometric Society}, pages
  571--599, 1988{\natexlab{b}}.

\bibitem[McKelvey and Palfrey(1995)]{ref:McKelvey95}
R.~McKelvey and T.~Palfrey.
\newblock Quantal response equilibria for normal form games.
\newblock \emph{Games and Economic Behavior}, 10:\penalty0 6--38, 1995.

\bibitem[McMahan(2011)]{mcmahan2011follow}
B.~McMahan.
\newblock Follow-the-regularized-leader and mirror descent: {E}quivalence
  theorems and $l_1$ regularization.
\newblock In \emph{International Conference on Artificial Intelligence and
  Statistics}, pages 525--533, 2011.

\bibitem[Milgrom and Roberts(1991)]{ref:Milgrom91}
P.~Milgrom and J.~Roberts.
\newblock Adaptive and sophisticated learning in normal form games.
\newblock \emph{Games and Economic Behavior}, 3:\penalty0 82--100, 1991.

\bibitem[Miyasawa(1961)]{ref:Miyasawa61}
K.~Miyasawa.
\newblock On the convergence of the learning process in a 2x2 non-zero-sum
  game.
\newblock \emph{Economic Research Program, Princeton University, Research
  Memorandum}, 33, 1961.

\bibitem[Monderer and Sela(1996)]{ref:Monderer96a}
D.~Monderer and A.~Sela.
\newblock A 2x2 game without the fictitious play property.
\newblock \emph{Games and Economic Behavior}, 14:\penalty0 144--148, 1996.

\bibitem[Monderer and Shapley(1996{\natexlab{a}})]{ref:Monderer96b}
D.~Monderer and L.~Shapley.
\newblock Fictitious play property for games with identical interests.
\newblock \emph{Games and Economic Behavior}, 68:\penalty0 258--265,
  1996{\natexlab{a}}.

\bibitem[Monderer and Shapley(1996{\natexlab{b}})]{ref:Monderer96c}
D.~Monderer and L.~Shapley.
\newblock Potential games.
\newblock \emph{Games and Economic Behavior}, 14:\penalty0 124--143,
  1996{\natexlab{b}}.

\bibitem[Nagel(1995)]{ref:Nagel95}
R.~Nagel.
\newblock Unraveling in guessing games: {A}n experimental study.
\newblock \emph{American Economic Review}, 5:\penalty0 1313--1326, 1995.

\bibitem[Perkins and Leslie(2012)]{ref:Perkins12}
S.~Perkins and D.~S. Leslie.
\newblock Asynchronous stochastic approximation with differential inclusions.
\newblock \emph{Stochastic Systems}, 2\penalty0 (2):\penalty0 409--446, 2012.

\bibitem[P{\'e}rolat et~al.(2015)P{\'e}rolat, Scherrer, Piot, and
  Pietquin]{perolat2015approximate}
J.~P{\'e}rolat, B.~Scherrer, B.~Piot, and O.~Pietquin.
\newblock Approximate dynamic programming for two-player zero-sum {M}arkov
  games.
\newblock In \emph{International Conference on Machine Learning}, pages
  1321--1329, 2015.

\bibitem[P{\'e}rolat et~al.(2017)P{\'e}rolat, Strub, Piot, and
  Pietquin]{pmlr-v54-perolat17a}
J.~P{\'e}rolat, F.~Strub, B.~Piot, and O.~Pietquin.
\newblock {Learning {N}ash Equilibrium for General-Sum {M}arkov Games from
  Batch Data}.
\newblock In \emph{International Conference on Artificial Intelligence and
  Statistics}, pages 232--241, 2017.

\bibitem[P{\'e}rolat et~al.(2018)P{\'e}rolat, Piot, and
  Pietquin]{perolat2018actor}
J.~P{\'e}rolat, B.~Piot, and O.~Pietquin.
\newblock Actor-critic fictitious play in simultaneous move multistage games.
\newblock In \emph{International Conference on Artificial Intelligence and
  Statistics}, pages 919--928, 2018.

\bibitem[Poljak and Tsypkin(1973)]{ref:Polyak73}
B.~T. Poljak and Y.~Z. Tsypkin.
\newblock Pseudogradient adaptation and training algorithms.
\newblock \emph{Automation and Remote Control}, 12:\penalty0 83--94, 1973.

\bibitem[Robinson(1951)]{robinson1951iterative}
J.~Robinson.
\newblock An iterative method of solving a game.
\newblock \emph{Annals of Mathematics}, pages 296--301, 1951.

\bibitem[Roughgarden(2009)]{roughgarden2009intrinsic}
T.~Roughgarden.
\newblock Intrinsic robustness of the price of anarchy.
\newblock In \emph{ACM Symposium on Theory of Computing}, pages 513--522, 2009.

\bibitem[Roughgarden(2010)]{roughgarden2010algorithmic}
T.~Roughgarden.
\newblock Algorithmic game theory.
\newblock \emph{Communications of the ACM}, 53\penalty0 (7):\penalty0 78--86,
  2010.

\bibitem[Sandholm(2001)]{ref:Sandholm01}
W.~H. Sandholm.
\newblock Preference evolution, two-speed dynamics, and rapid social change.
\newblock \emph{Review of Economic Dynamics}, 4\penalty0 (3):\penalty0
  637--679, 2001.

\bibitem[Sayin et~al.(2020)Sayin, Parise, and Ozdaglar]{ref:Sayin20}
M.~O. Sayin, F.~Parise, and A.~Ozdaglar.
\newblock Fictitious play in zero-sum stochastic games.
\newblock \emph{arXiv:2010.04223}, 2020.

\bibitem[Sayin et~al.(2021)Sayin, Zhang, Leslie, Ba\c{s}ar, and
  Ozdaglar]{ref:Sayin21}
M.~O. Sayin, K.~Zhang, D.~S. Leslie, T.~Ba\c{s}ar, and A.~Ozdaglar.
\newblock Decentralized {Q}-learning in zero-sum markov games.
\newblock In \emph{Thirty-fifth Conference on Neural Information Processing
  Systems}, 2021.

\bibitem[Sela(1999)]{ref:Sela99}
A.~Sela.
\newblock Fictitious play in ``one-against-all'' multi-player games.
\newblock \emph{Economic Theory}, 14:\penalty0 635--651, 1999.

\bibitem[Shah et~al.(2020)Shah, Somani, Xie, and Xu]{shah2020reinforcement}
D.~Shah, V.~Somani, Q.~Xie, and Z.~Xu.
\newblock On reinforcement learning for turn-based zero-sum {M}arkov games.
\newblock \emph{arXiv preprint arXiv:2002.10620}, 2020.

\bibitem[Shalev-Shwartz et~al.(2016)Shalev-Shwartz, Shammah, and
  Shashua]{shalev2016safe}
S.~Shalev-Shwartz, S.~Shammah, and A.~Shashua.
\newblock Safe, multi-agent, reinforcement learning for autonomous driving.
\newblock \emph{arXiv preprint arXiv:1610.03295}, 2016.

\bibitem[Shalev-Shwartz et~al.(2011)]{shalev2011online}
S.~Shalev-Shwartz et~al.
\newblock Online learning and online convex optimization.
\newblock \emph{Foundations and Trends in Machine Learning}, 4\penalty0
  (2):\penalty0 107--194, 2011.

\bibitem[Shapley(1953)]{ref:Shapley53}
L.~S. Shapley.
\newblock Stochastic games.
\newblock \emph{Proceedings of National Academy of Science USA}, 39\penalty0
  (10):\penalty0 1095--1100, 1953.

\bibitem[Shapley(1964)]{shapley1964some}
L.~S. Shapley.
\newblock Some topics in two-person games.
\newblock \emph{Advances in Game Theory}, 52:\penalty0 1--29, 1964.

\bibitem[Shoham and Leyton-Brown(2008)]{ref:MAS}
Y.~Shoham and K.~Leyton-Brown.
\newblock \emph{Multiagent Systems: Algorithmic, Game-theoretic, and Logical
  Foundations}.
\newblock Cambridge University Press, 2008.

\bibitem[Sidford et~al.(2020)Sidford, Wang, Yang, and Ye]{sidford2019solving}
A.~Sidford, M.~Wang, L.~Yang, and Y.~Ye.
\newblock Solving discounted stochastic two-player games with near-optimal time
  and sample complexity.
\newblock In \emph{International Conference on Artificial Intelligence and
  Statistics}, pages 2992--3002, 2020.

\bibitem[Silver et~al.(2016)Silver, Huang, Maddison, Guez, Sifre, Van
  Den~Driessche, Schrittwieser, Antonoglou, Panneershelvam, Lanctot,
  et~al.]{silver2016mastering}
D.~Silver, A.~Huang, C.~J. Maddison, A.~Guez, L.~Sifre, G.~Van Den~Driessche,
  J.~Schrittwieser, I.~Antonoglou, V.~Panneershelvam, M.~Lanctot, et~al.
\newblock Mastering the game of \relax{G}o with deep neural networks and tree
  search.
\newblock \emph{Nature}, 529\penalty0 (7587):\penalty0 484--489, 2016.

\bibitem[Silver et~al.(2017)Silver, Schrittwieser, Simonyan, Antonoglou, Huang,
  Guez, Hubert, Baker, Lai, Bolton, et~al.]{silver2017mastering}
D.~Silver, J.~Schrittwieser, K.~Simonyan, I.~Antonoglou, A.~Huang, A.~Guez,
  T.~Hubert, L.~Baker, M.~Lai, A.~Bolton, et~al.
\newblock Mastering the game of \relax{G}o without human knowledge.
\newblock \emph{Nature}, 550\penalty0 (7676):\penalty0 354--359, 2017.

\bibitem[Syrgkanis and Tardos(2013)]{syrgkanis2013composable}
V.~Syrgkanis and E.~Tardos.
\newblock Composable and efficient mechanisms.
\newblock In \emph{ACM Symposium on Theory of Computing}, pages 211--220, 2013.

\bibitem[Szepesv{\'a}ri and Littman(1999)]{szepesvari1999unified}
C.~Szepesv{\'a}ri and M.~L. Littman.
\newblock A unified analysis of value-function-based reinforcement-learning
  algorithms.
\newblock \emph{Neural Computation}, 11\penalty0 (8):\penalty0 2017--2060,
  1999.

\bibitem[Takahashi(1964)]{ref:Takahashi64}
M.~Takahashi.
\newblock Equilibrium points of stochastic non-cooperative n-person games.
\newblock \emph{Journal of Science Hiroshima University Series A-I},
  28:\penalty0 95--99, 1964.

\bibitem[Tan(1993)]{tan1993multi}
M.~Tan.
\newblock Multi-agent reinforcement learning: {I}ndependent vs. cooperative
  agents.
\newblock In \emph{International Conference on Machine Learning}, pages
  330--337, 1993.

\bibitem[Tian et~al.(2021)Tian, Wang, Yu, and Sra]{tian2020provably}
Y.~Tian, Y.~Wang, T.~Yu, and S.~Sra.
\newblock Online learning in unknown {M}arkov games.
\newblock In \emph{International Conference on Machine Learning}, pages
  10279--10288, 2021.

\bibitem[Tsitsiklis(1994)]{ref:Tsitsiklis94}
J.~N. Tsitsiklis.
\newblock Asynchronous stochastic approximation and {Q}-learning.
\newblock \emph{Machine Learning}, 16:\penalty0 185--202, 1994.

\bibitem[{Van der Genugten}(2000)]{ref:Genugten00}
B.~{Van der Genugten}.
\newblock A weakened form of fictitious play in two-person zero-sum games.
\newblock \emph{International Game Theory Review}, 2\penalty0 (4):\penalty0
  307--328, 2000.

\bibitem[Watkins and Dayan(1992)]{ref:Watkins92}
C.~J. C.~H. Watkins and P.~Dayan.
\newblock Q-learning.
\newblock \emph{Machine Learning}, 8\penalty0 (3):\penalty0 279--292, 1992.

\bibitem[Wei et~al.(2017)Wei, Hong, and Lu]{wei2017online}
C.-Y. Wei, Y.-T. Hong, and C.-J. Lu.
\newblock Online reinforcement learning in stochastic games.
\newblock In \emph{Advances in Neural Information Processing Systems}, pages
  4987--4997, 2017.

\bibitem[Wei et~al.(2021)Wei, Lee, Zhang, and Luo]{wei2021last}
C.-Y. Wei, C.-W. Lee, M.~Zhang, and H.~Luo.
\newblock Last-iterate convergence of decentralized optimistic gradient
  descent/ascent in infinite-horizon competitive {Markov} games.
\newblock In \emph{Conference on Learning Theory}, 2021.

\bibitem[Xie et~al.(2020)Xie, Chen, Wang, and Yang]{xie2020learning}
Q.~Xie, Y.~Chen, Z.~Wang, and Z.~Yang.
\newblock Learning zero-sum simultaneous-move {M}arkov games using function
  approximation and correlated equilibrium.
\newblock In \emph{Conference on Learning Theory}, pages 3674--3682, 2020.

\bibitem[Yang et~al.(2020)Yang, Li, and Peng]{yang2020multi}
Y.~Yang, J.~Li, and L.~Peng.
\newblock Multi-robot path planning based on a deep reinforcement learning
  {DQN} algorithm.
\newblock \emph{CAAI Transactions on Intelligence Technology}, 5\penalty0
  (3):\penalty0 177--183, 2020.

\bibitem[Zhang et~al.(2019)Zhang, Yang, and Ba\c{s}ar]{zhang2019policyb}
K.~Zhang, Z.~Yang, and T.~Ba\c{s}ar.
\newblock Policy optimization provably converges to {Nash} equilibria in
  zero-sum linear quadratic games.
\newblock In \emph{Advances in Neural Information Processing Systems}, pages
  11598--11610, 2019.

\bibitem[Zhang et~al.(2020)Zhang, Kakade, Ba\c{s}ar, and Yang]{zhang2020model}
K.~Zhang, S.~Kakade, T.~Ba\c{s}ar, and L.~Yang.
\newblock Model-based multi-agent {RL} in zero-sum markov games with
  near-optimal sample complexity.
\newblock \emph{Advances in Neural Information Processing Systems}, 33, 2020.

\bibitem[Zhang et~al.(2021)Zhang, Yang, and Ba\c{s}ar]{ref:Zhang21}
K.~Zhang, Z.~Yang, and T.~Ba\c{s}ar.
\newblock Multi-agent reinforcement learning: {A} selective overview of
  theories and algorithms.
\newblock In \emph{Handbook of Reinforcement Learning and Control}, Studies in
  Systems, Decision and Control. Springer, 2021.

\bibitem[Zhao et~al.(2021)Zhao, Tian, Lee, and Du]{zhao2021provably}
Y.~Zhao, Y.~Tian, J.~D. Lee, and S.~S. Du.
\newblock Provably efficient policy gradient methods for two-player zero-sum
  {M}arkov games.
\newblock \emph{arXiv preprint arXiv:2102.08903}, 2021.

\bibitem[Zinkevich(2003)]{zinkevich2003online}
M.~Zinkevich.
\newblock Online convex programming and generalized infinitesimal gradient
  ascent.
\newblock In \emph{International Conference on Machine Learning}, pages
  928--936, 2003.

\end{thebibliography}
\end{spacing}

\end{document}